\documentclass[lettersize,journal]{IEEEtran}
\usepackage{cite}
\usepackage{graphicx}
\usepackage{verbatim}
\usepackage{float}
\usepackage{stfloats}
\usepackage{url}
\usepackage{bm}
\usepackage{array}
\usepackage[caption=false,font=normalsize,labelfont=sf,textfont=sf]{subfig}
\usepackage{textcomp}
\usepackage{amsmath}
\usepackage{amsthm} 
\usepackage{amsxtra}
\usepackage{mathrsfs} 
\usepackage{amssymb} 
\usepackage{amsfonts} 
\usepackage{algorithm}
\usepackage{algorithmic}
\usepackage{setspace}
\usepackage{multirow}
\usepackage[T1]{fontenc}
\usepackage{graphicx}
\newtheorem{definition}{Definition}
\newtheorem{theorem}{Theorem}

\newtheorem{lemma}{Lemma}

\def\BibTeX{{\rm B\kern-.05em{\sc i\kern-.025em b}\kern-.08em
    T\kern-.1667em\smaller.7ex\hbox{E}\kern-.125emX}}
\usepackage{balance}
\begin{document}
	\title{A Construction of Evolving $3$-threshold Secret Sharing Scheme with Perfect Security and Smaller Share Size}
	
	\author{Qi Cheng, Hongru Cao, and Sian-Jheng Lin,~\IEEEmembership{Member,~IEEE}
	\thanks{This work was supported by the National Natural Science Foundation of China under Grant 62071446. (Corresponding author: Sian-Jheng Lin.)}
	\thanks{Qi Cheng is with the School of Information and Artificial Intelligence, Anhui Agricultural University, Hefei 230036, China (e-mail:chengqi@ahau.edu.cn)}
	\thanks{Hongru Cao and Sian-Jheng Lin are with the CAS Key Laboratory of Electromagnetic Space Information, School of Cyber Science and Technology, University of Science and Technology of China, Hefei 230027, China (e-mail: chrkeith@mail.ustc.edu.cn; sjlin@ustc.edu.cn).}
	}	
	\maketitle
\thispagestyle{empty}

\begin{abstract}
	The evolving $k$-threshold secret sharing scheme allows the dealer to distribute the secret to many participants such that only no less than $k$ shares together can restore the secret. In contrast to the conventional secret sharing scheme, the evolving scheme allows the number of participants to be uncertain and even ever-growing. In this paper, we consider the evolving secret sharing scheme with $k=3$. First, we point out that the prior approach has risks in the security. To solve this issue, we then propose a new evolving $3$-threshold scheme with perfect security. Given a $\ell$-bit secret, the $t$-th share of the proposed scheme has $\lceil\log_2 t\rceil +O({\lceil \log_4 \log_2 t\rceil}^2)+\log_2 p(2\lceil \log_4 \log_2 t\rceil-1)$ bits, where $p$ is a prime. Compared with the prior result $2 \lfloor\log_2 t\rfloor+O(\lfloor\log_2 t\rfloor)+\ell$, the proposed scheme reduces the leading constant from $2$ to $1$. 
	Finally, we propose a conventional $3$-threshold secret sharing scheme over a finite field. Based on this model of the revised scheme and the proposed conventional $3$-threshold scheme, we present a brand-new and more concise evolving $3$-threshold secret sharing scheme. %Lin: need revise
\end{abstract}


\section{Introduction}
\IEEEPARstart{S}{ecret} sharing scheme is a protocol dividing a secret into many shares, that are distributed to corresponding participants so that only a certain number of shares can recover the secret. Therefore, when using secret sharing schemes to transmit secret information, the security of the information can be ensured by resisting multi-party intrusions. The conventional $(k, n)$-threshold secret sharing scheme divides the secret into $n$ shares such that any $k$ out of $n$ shares can reconstruct the secret, and any $k-1$ shares cannot obtain any information about the secret. In 1979, Shamir~\cite{shamir1979share} and Blakley~\cite{blakley1979safeguarding} independently proposed the $(k,n)$ secret sharing scheme. In Shamir's scheme, the dealer randomly choose a polynomial in $F_p[x]$ with the degree less than $k$ and the constant term of this polynomial as the secret, where $p>n$. Consequently, the $n$ shares are obtained by evaluate the polynomial at $n$ distinct points. Therefore, each share is an element of $F_p$, that requires $\lceil\lg p\rceil$ bits, where $\lg$ denotes the binary logarithm. The secret can be reconstructed by any $k$ shares with using Lagrange interpolations.
With subsequent developments, many schemes~\cite{fuyou2014randomized,harn2016realizing,harn2010strong,pang2005new,yang2004t} have been proposed in recent decades.

For conventional secret sharing schemes, the maximum number of participants $n$ is required to be known before generating the shares. However, when the dealer allows that new participants can join the secret-sharing group, we cannot determine the maximum of $n$, and the conventional schemes cannot be applied. Therefore, an evolving scheme~\cite{csirmaz2012line} was introduced. Then Komargodski et al.~\cite{komargodski2016share,komargodski2017share} proposed evolving $k$-threshold secret sharing scheme with perfect security for infinitely many participants. In an evolving $k$-threshold scheme, the secret can be reconstructed by any $k$ shares (correctness), and any $k-1$ or fewer shares cannot obtain any information about the secret (security). Komargodski et al.'s scheme left the possibility of constructing an efficient scheme for the dynamic threshold access structure, then Komargodski et al.~\cite{komargodski2017evolving} gave the construction by using the algebraic manipulation detection codes, which resolves this open
problem mentioned in \cite{komargodski2017share}. Besides, they also proposed the method to transform any evolving $k$-threshold secret sharing scheme into a robust scheme, which still can reconstruct the secret when some participants submit incorrect shares. Moreover, Dutta et al.~\cite{dutta2019secret} proposed an evolving secret sharing scheme when considering the compartmental and hierarchical multi-level access structures.

There exist two important factors that need to be considered, when constructing a generally evolving secret sharing scheme. The first one is to ensure correctness and security, the other is to achieve the lowest share size possible. The correctness and security can guarantee information privacy when performing multi-party operations. The share size can measure the performance of the proposed scheme, specifically, a smaller share size can improve the effectiveness of communication and storage. However, it is not easy to construct the evolving secret sharing schemes having a smaller share size satisfying perfect correctness and security. Thus, it has always been one of the research focuses in the evolving secret sharing field.   

When $k=2$, based on prefix coding of the positive integer, Komargodski et al.~\cite{komargodski2017share} first proposed an evolving $2$-threshold secret sharing scheme and proved the corresponding correctness and security. Besides, they also analyzed the corresponding share size. For an $\ell$-bit secret, it shows that the $t$-th share size is no more than $\lg t+(\ell+1)\lg{\lg t}+4\ell+1$ bits. The result also implies that the scheme is optimal for a $1$-bit secret. Then, D'Arco et al.~\cite{d2018equivalence} showed the equivalence between evolving $2$-threshold schemes and binary prefix codes, which implies that when a set of prefix codes is given, there must be a corresponding construction method for the $2$-threshold scheme, and vice versa. Then, Okamura et al.~\cite{okamura2020new} proposed a new construction of the evolving $2$-threshold scheme based on the binary prefix code for a given secret with arbitrary length. Besides, they also gave the construction of an evolving $2$-threshold scheme based on a $D$-ary prefix coding with $D\geq 3$. To obtain a smaller share size, Cheng et al.~\cite{cheng2024construction} proposed an evolving $2$-threshold secret sharing scheme for an $\ell$-bit secret over a polynomial ring based on prefix coding, which is proven to have perfect correctness and secrecy. In addition, they also analyzed the corresponding share size. The $t$-th share size is $\ell_t+\ell-1$ bits, where $\ell_t$ represents the bit length for encoding $t$. The result showed that compared with the scheme proposed in~\cite{komargodski2017share}, the proposed scheme can achieve a smaller size when using $\gamma$ coding or $\delta$ coding~\cite{elias1975universal}.
%The proposed scheme can provided a unified mathematical decryption for prior evolving $2$-threshold secret sharing schemes. The authors aslo applied the construction to other complicated binary prefix codes and analyzed the share size. It shows that .

When $k\geq 3$, Komargodski et al.~\cite{komargodski2017share}  proposed a brand-new evolving $k$-threshold secret sharing scheme for an $\ell$-bit secret. They~\cite{komargodski2017share} also proved correctness and secrecy and analyzed the corresponding share size. The results of share size showed that the $t$-th share size is $(k-1)\lg t+6k^4\ell\lg{\lg t}\cdot\lg{\lg {\lg t}}+ 7k^4\ell\lg k$, which is not optimal. %Lin: how to know it is not optimal?
On the other hand, the proposed scheme was constructed based on the idea of distributing shares on a generational basis by using a general evolving scheme once and Shamir's scheme multiple times, which led to it becoming increasingly complex and even not easy to construct as $k$ increases. To construct the simpler scheme with a smaller share size, based on the prefix coding, Cheng et al.~\cite{cheng2024construction} proposed an evolving $k$-threshold secret sharing scheme over a polynomial ring with perfect correctness and secrecy. Compared with the scheme in~\cite{komargodski2017share}, the proposed scheme is more concise. By analyzing, the $t$-th share size is $(k-1)(\ell_t-1)+\ell$ bits. Specifically, when using $\delta$ code~\cite{elias1975universal} as the prefix code, the share size is given by $(k-1)\lfloor\lg t\rfloor+2(k-1)\lfloor\lg ({\lfloor\lg t\rfloor+1}) \rfloor+\ell$, which is smaller than the result in~\cite{komargodski2017share}. However, the result of share size is still not optimal. %Lin: how to know it is not optimal?

Therefore, it is still an unresolved issue to construct the optimal evolving $k$-threshold secret sharing scheme. It was found that in the proposed two evolving $k$-threshold schemes~\cite{komargodski2017share,cheng2024construction}, the share size of the $t$-th party is both $(k-1)\lg t+ poly (k)o(\lg t)$ by a unified mathematical expression and the most important term is both $(k-1)\lg t$, where $poly (k)$ represents a polynomial of $k$ and $o(\lg t)$ represents a high-order infinitesimal about $\lg t$. A natural idea is whether to propose a new scheme for the dependency of the most important term from $k$ can be avoided or reduced. To solve the issue, D'Arco et al.~\cite{DARCO2021149} proposed a new evolving $3$-threshold scheme based on the Chinese Remainder Theorem to reduce the share size. They made the share size of the $t$-th participant is ${\frac{4}{3}\lg t+c(\log_4{\lg t})^2+\lg p(\log_4{\lg t})}$, which achieve the smaller share size. However, in terms of our analysis, the author obtained incorrect results in the proof of the security, therefore the scheme does not meet the perfect security.

To this end, we first carefully analyze the security proof of the scheme~\cite{DARCO2021149} and identify the underlying issues. Then, based on the scheme~\cite{DARCO2021149}, we attempt to propose a revised scheme model satisfying perfect security, and the corresponding share size is no more than the best result in ~\cite{cheng2024construction}. By analyzing the model of the revised scheme, we would like to propose a more concise construction method that also satisfies perfect security with the corresponding share size as small as possible. The main contributions of this paper are enumerated as follows. 
\begin{itemize}
	\item[$\bullet$] The revised evolving $3$-threshold secret sharing scheme with perfect security is proposed, which solves the underlying issue of the scheme in~\cite{DARCO2021149}.
	
	\item[$\bullet$] The $t$-th share size of the proposed revised scheme is $\lg t+O({\lceil \log_4 \lg t\rceil}^2)+\lg p(2\lceil \log_4 \lg t\rceil-1)$, that improves the prior results~\cite{komargodski2017share,cheng2024construction} by reducing the leading term from $2\lg t$ into $\lg t$.
	
	\item[$\bullet$] A new construction of the conventional $3$-threshold secret shaing scheme over a finite field $F_{2^{\ell m}}$ with perfect security, is proposed. The scheme is interesting and is different from Shamir's scheme. %Lin: this should be revised
	
	\item[$\bullet$] Based on the model of the proposed revised scheme, a new construction of the evolving $3$-threshold scheme is proposed. Compared with the revised scheme, the proposed new scheme is more concise while maintaining the same size of share length. 	
\end{itemize}

The rest of the paper is organized as follows. In Section~\ref{sec2}, we introduce the traditional secret sharing scheme and evolving secret sharing scheme. In Section~\ref{sec3}, we review the evolving $3$-threshold secret sharing scheme proposed in~\cite{DARCO2021149}, and analyze the underlying issues. In Section~\ref{sec4}, based on the proposed scheme in~\cite{DARCO2021149}, we propose a revised scheme model with perfect security and analyze the corresponding share size. Specifically, in Section~\ref{sec5}, we propose a new construction of an evolving $3$-threshold secret sharing scheme. Finally, Section~\ref{sec6} concludes the work and discusses the unresolved issues.

\section{Models and Notations}\label{sec2}
%Let $\mathbb{N}^+$ denote the set of positive integers. Let $[n]=\{1,2,\cdots,n\}$ for $n\in \mathbb{N}^+$. Let $|A|$ denote the cardinality of the set $A$. For any $p\in \mathbb{N}^+$ being a prime or a power of the prime, let $F_p$ denote the finite field with $p$ elements. Let $F_p[x]=\{\sum_{j=0}^{N}a_jx^{j}|a_j\in F_p\}$ as the polynomial ring, where $N$ is a finite positive integer. For any $f(x), g(x)\in F_p[x]$, if $g(x)$ divides $f(x)$, it can be expressed as $g(x)\mid f(x)$. The result of $g(x)$ dividing $f(x)$ is denoted as quotient, which is written as $\frac{f(x)}{g(x)}$. For any $i,a\in \mathbb{N}^+$, $f(x)\in F_p[x]$, if $x^i\mid f(x)$ with any $i\leq a$ and $x^{a+1}\nmid f(x)$, then $a$ be denoted as the largest integer of $f(x)$. For any $d\in \mathbb{N}^+$, let $F_p[x]/x^d$ be a quotient ring, which is composed of polynomials $f(x)\in F_p[x]$ with the degree less than $d$. For any $f(x)\in F_p[x]/x^d$, if there exists a polynomial $g(x)\in F_p[x]/x^d$ such that $f(x)g(x)=1 \in F_p[x]/x^d$, then $f(x)$ is called to be reversible, which be respresented as $f^{-1}(x)$. Let $F_p[[x]]=\{\sum_{j=0}^{\infty}a_jx^{j}|a_j\in F_p\}$.

The following gives the notations used in this paper. Let $\mathbb{N}^+$ denote the set of positive integers. Let $|A|$ denote the cardinality of the set $A$. For any $p\in \mathbb{N}^+$ being a prime, let $F_p$ denote the finite field with $p$ elements.  
\subsection{Secret Sharing Scheme} 
Denote $\mathcal{P}_n=\{P_1,P_2,\cdots,P_n\}$as the set of $n$ participants. $2^{\mathcal{P}_n}$ represents the power set of $\mathcal{P}_n$. Let $\mathcal{A} \subseteq 2^{\mathcal{P}_n}$, if for arbitrary $A\in \mathcal{A}$ and $A\subseteq C\in 2^{\mathcal{P}_n}$, it holds that $C\in \mathcal{A}$, then the  collection $\mathcal{A}$ is monotone. The access structure is defined as follows.
\begin{definition}\label{def1}
	$\mathcal{A} \subseteq 2^{\mathcal{P}_n}$ is called an access structure if $\mathcal{A}$ is a monotone collection of non-empty subsets. The subset in $\mathcal{A}$ is called qualified, and the subset in $2^{\mathcal{P}_n}\setminus\mathcal{A}$ is called unqualified. 
\end{definition}

\begin{definition}\label{def2}
	For $k,n \in \mathbb{N}^+$ with $k\leq n$, 
	the $(k,n)$ threshold access structure $\mathcal{A}$ is an access structure, and the size of each element in $\mathcal{A}$ is no less than $k$, i.e
	\begin{equation*}
		\mathcal{A}=\{A\in 2^{\mathcal{P}_n}||A|\geq k\}.	
	\end{equation*}
\end{definition}

Actually, a $(k,n)$-threshold secret sharing scheme requires a set of $n$ participants and a $(k,n)$ access structure $\mathcal{A}$, and a given secret $s\in S$, where $S$ is the domain of the secret. Then the dealer distributes the share to every participant such that the shares of any subset in $\mathcal{A}$ can correctly recover the secret, while the shares of any subset not in $\mathcal{A}$ cannot gain any information about the secret. 

Let ${Z}^{(s)}_i$ denote the share of the $i$-th participant and $B({Z}^{(s)}_i)$ represent the bit length of ${Z}^{(s)}_i$ for $1\leq i\leq n$. The $(k,n)$-threshold secret sharing scheme is composed of a pair of algorithms $(\mathcal{E},\mathcal{R})$, where $\mathcal{E}$ is used to encode the secret $s$ into shares, and $\mathcal{R}$ is used to restore the secret from a subset of shares $H\in 2^{\mathcal{P}_n}$. The following requirements shall be satisfied.
\begin{itemize}
	\item[$\bullet$] Correctness: For any qualified set $A\in \mathcal{A}$, the algorithm $\mathcal{R}$ can correctly recover $s$ from the shares of the participants in $A$, that is
	\begin{equation}
		P[\mathcal{R}(\{Z^{(s)}_i\}_{P_i\in A},A)=s]=1.
	\end{equation}
	\item[$\bullet$] Security: For any unqualified set $C\in 2^{\mathcal{P}_n}\setminus\mathcal{A}$, there is no information about $s$ leaking to the participants in $C$.
\end{itemize} %该到这里
In particular, the following conclusion is usually used to verify the security of the secret sharing schemes.
\begin{lemma}[\cite{komargodski2017share}]\label{lem1}
	Let $s_0, s_1\in S$ be two different secrets. The scheme is secure if for arbitrary $C\in2^{\mathcal{P}_n}\setminus\mathcal{A}$ and arbitrary $\{z_i\}_{P_i\in C}$,  the following two probabilities are equal, i.e.
	\begin{equation}
		P(\{Z^{(s_0)}_i=z_i\}_{P_i\in C})=P(\{Z^{(s_1)}_i=z_i\}_{P_i\in C}),
	\end{equation}
	where $z_i$ is the possible share assigned to the participant $P_i$, $\{Z^{(s_0)}_i\}_{P_i\in C}$ and $\{Z^{(s_1)}_i\}_{P_i\in C}$ are the corresponding shares distributed to the participants of $C$ under two different secrets $s_0$ and $s_1$, respectively,
\end{lemma}

Shamir~\cite{shamir1979share} first proposed a construction for the $(k, n)$-threshold secret sharing scheme. For an $\ell$-bit secret $s$, the scheme uses the polynomial over $F_p$ for $p\geq n$. The share size $Z^{(s)}_i$ satisfies $B(Z^{(s)}_i)\geq \max\{\ell,\lg p\}$ for any $i\in [n]$ in Shamir's scheme.

\subsection{Evolving Secret Sharing Scheme}
When the maximum number of $n$ cannot be determined in advance and can even be infinite, conventional secret sharing schemes cannot be applied directly. In this case, evolving secret sharing schemes are developed. We denote $\mathcal P=\{P_1, P_2, \cdots, P_n, \cdots\}$ as the set of participants, where $ \mathcal P$ is possibly infinite. Then, we give some definitions of the evolving secret sharing scheme. 

\begin{definition}\label{def3}
	If $\mathcal{A} \subseteq 2^{\mathcal{P}}$ is monotone, and for any $t\in \mathbb{N}^+$, $\mathcal{A}_t \colon=\mathcal{A}\cap2^{\{P_1,P_2,\cdots,P_t\}}$ is an access structure, then $\mathcal{A}$ is called an evolving access structure.
\end{definition}

\begin{definition}\label{def4}
	For $k \in \mathbb{N}^+$, 
	the evolving $k$-threshold access structure $\mathcal{A}$ is an evolving access structure whose elements are with sizes no less than $k$, i.e
	\begin{equation*}
		\mathcal{A}=\{A\in 2^{\mathcal{P}}||A|\geq k\}.	
	\end{equation*}
\end{definition}

Similarly, an evolving $k$-threshold secret sharing scheme requires a secret $s\in S$, a set of participants and an evolving $k$-threshold access structure $\mathcal{A}$. In the scheme, the secret can be recovered from any $k$ out of infinitely many participants, and any $k-1$ shares cannot deduce any information about the secret. An evolving $k$-threshold secret sharing scheme also includes a pair of algorithms $(\mathcal{E},\mathcal{R})$, which satisfy the following requirements.  
\begin{itemize}
	\item[$\bullet$]Composition: For any $j\in \mathbb{N}^+$, the share $Z^{(s)}_j$ of $j$-th participant is constructed by the previous $j-1$ shares $\{Z^{(s)}_i\}_{i=1}^{j-1}$ using the algorithm $\mathcal{E}$, i.e.  
	\begin{equation}
		Z^{(s)}_j=\mathcal{E}(s,\{Z^{(s)}_i\}_{i=1}^{j-1}).
	\end{equation}
	
	\item[$\bullet$]Correctness: For any $t\in \mathbb{N}^+$, $A\in \mathcal{A}_t$, the algorithm $\mathcal{R}$ can correctly recover the secret $s$ from the shares of the participants in $A$.
	
	\item[$\bullet$]Secrecy: For any $t\in \mathbb{N}^+, C\in 2^{\mathcal{P}_t}\setminus\mathcal{A}_t$, there is no information about $s$ leaking to the participants in $C$.    
\end{itemize} 

%and obtained two main results, which are described as follows. 

%\begin{theorem}
%	For $\ell,t\in \mathbb{N}^+$, there exists an evolving $2$-threshold secret sharing scheme, where the size of the $t$-th share satisfies
%	\begin{equation}
%		B(Z^{(s)}_t)\leq \lg t+(\ell+1)\lg{\lg t}+4\ell+1.
%	\end{equation}
%\end{theorem}
%
%\begin{theorem}
%	For $\ell,k, t\in \mathbb{N}^+$, there exists an evolving $k$-threshold secret sharing scheme, where the size of the $t$-th share satisfies
%	\begin{equation}
%		B(Z^{(s)}_t)\leq (k-1)\lg t+6k^4\ell\lg{\lg t}\cdot\lg{\lg {\lg t}}+ 7k^4\ell\lg k.
%	\end{equation}
%\end{theorem}

\section{The Security Vulnerabilities of the Evolving $3$-threshold Secret Sharing Scheme Proposed in~\cite{DARCO2021149}}\label{sec3}
In 2017, Komargodski et al.~\cite{komargodski2016share,komargodski2017share} first constructed an evolving $k$-threshold secret sharing scheme for an ${\ell}$-bit secret $s$, and obtained a main result, which is described as follows.
\begin{theorem}\label{thm1}
	For $\ell,k, t\in \mathbb{N}^+$, there exists an evolving $k$-threshold secret sharing scheme, where the size of the $t$-th share satisfies
	\begin{equation}
		B(Z^{(s)}_t)\leq (k-1)\lg t+6k^4\ell\lg{\lg t}\cdot\lg{\lg {\lg t}}+ 7k^4\ell\lg k.
	\end{equation}
\end{theorem}

However, the share size of the scheme~\cite{komargodski2016share,komargodski2017share} is not optimal for any $k$, where $k\geq 3$. To obtain a smaller share size, D'Arco et al.~\cite{DARCO2021149} proposed a new evolving $3$-threshold scheme. 

\subsection{Proposed Scheme in \cite{DARCO2021149}}\label{sub3.1}
Let's review the scheme proposed in~\cite{DARCO2021149}. Denote $\mathcal P=\{P_1, P_2, \cdots, P_n, \cdots\}$ as the set of participants. Each participant is assigned to
a generation when it joins the scheme. The generations are growing in a doubly exponential size, as follows. For $i=1,2,\cdots$, the $i$-th generation denoted by $G^i$. Let $n_0=0$ and $n_i={2^{4^i}}$ for $i\geq 1$, $G^i$ begins when the $(n_{i-1}+1)$-th participant joins, and ends when the $n_i$-th participant joins. Therefore, the size of the $i$-th generation $G^i$, i.e. the number of participants of $G^i$ is $S(G^i)=n_i-n_{i-1}=2^{4^{i-1}}(2^4-1)$. The $t$-th participant belongs to the $g(t)$-th generation given by the membership function $g(t) = \lceil \log_4 \lg t\rceil$. 

Given an $\ell$-bit secret $s$, the idea of the scheme proposed in~\cite{DARCO2021149} is to use a general evolving $3$-threshold secret sharing scheme $\mathcal{S}^{\infty}$ among different generations, and combines to use a $(3, S(G^i)+2)$-threshold secret sharing scheme $\mathcal{S}^{i}$ among the $i$-th generation, which is based on the Chinese remainder theorem. The evolving scheme $\mathcal{S}^{\infty}$ constructs many shares, the number of shares is equal to the number of generations and may be infinite. %Each participant of the $i$-th generation is distributed by the same share ${sh}_{i}^{\infty}$ via the scheme $\mathcal{S}^{\infty}$. 
For the $i$-th generation with $S(G^i)$ participants, $(3,S(G^i)+2)$-threshold scheme $\mathcal{S}^{i}$ constructs $S(G^i)+2$ shares, denoted by ${sh}_{1}^{i}, {sh}_{2}^{i}, \cdots, {sh}_{S(G^i)}^{i}, {sh}_{F}^{i}, {sh}_{B}^{i}$.

Precisely, for the $t$-th participant, it belongs to the $g(t)=\lceil \log_4 \lg t\rceil$-th generation, and the size of the $g(t)$-th generation is $S(G^{g(t)})$. We denote the corresponding index of $t$-th participant among $g(t)$-th generation as $h(t)$, i.e. $1\leq h(t)\leq S(G^{g(t)})$. The whole shares of the $t$-th participant are composed of several pieces as follows.

P1. The share ${sh}_{g(t)}^{\infty}$, is distributed by $\mathcal{S}^{\infty}$ among different generations.

P2. There are $g(t)-1$ forward shares. For $j=1,2,\cdots,g(t)-1$, each forward share ${sh}_{F}^{j}$ is one of shares constructed by the scheme $\mathcal{S}^{j}$.

P3. One of the shares ${sh}_{1}^{g(t)}, {sh}_{2}^{g(t)}, \cdots, {sh}_{S(G^{g(t)})}^{g(t)}$, ${sh}_{F}^{g(t)}, {sh}_{B}^{g(t)}$ except ${sh}_{F}^{g(t)}$ and ${sh}_{B}^{g(t)}$, we denoted by ${sh}_{h(t)}^{g(t)}$ for $1\leq h(t)\leq S(G^{g(t)})$. These $S(G^{g(t)})+2$ shares are constructed by the scheme $\mathcal{S}^{g(t)}$. 

P4. There are $g(t)-1$ copies of the backward shares. For $1\leq j\leq g(t)-1$, each backward share masked with a different random string is described as ${SR}^{j}\oplus {sh}_{B}^{g(t)}$. The share ${sh}_{B}^{g(t)}$ is constructed by the scheme $\mathcal{S}^{g(t)}$.

P5. A random string ${SR}^{g(t)}$.
%The whole shares of the $t$-th participant are shown in Fig.~\ref{fig3.1}. 
Actually, for any participant in the first generation $G^1$, the corresponding shares are composed of four parts $P1$, $P3$, $P4$, and $P5$ without $P2$. 
%\begin{figure}[!t]
%	\centering
%	\includegraphics[width=2.75in]{fig3.1.pdf}
%	\caption{The whole shares of the $t$-th participant.}
%	\label{fig3.1}
%\end{figure}

\subsection{The Corresponding Correctness}\label{sec3.2}
For any $A \in\mathcal{A}$, then $|A|\geq 3$. We need to show that the $\ell$-bit secret $s$ can be correctly reconstructed by the shares of the participants in $A$. Since the cases of $|A|\geq 3$ include the case of $|A|=3$, we only prove the case of $|A|=3$.  

Without loss of generality, we denote the three participants of $A$ as $P_i$, $P_j$, and $P_k$ with $i< j< k$. There exist the following four cases.

\noindent\textbf{Case 1} $P_i$, $P_j$ and $P_k$ belong to different generations.

\noindent\textbf{Case 2} $P_i$, $P_j$ and $P_k$ belong to the same generation.

\noindent\textbf{Case 3} $P_i$ and $P_j$ belong to the same generation $G^{i_1}$, and $P_k$ belongs to another generation $G^{i_2}$, where $i_1<i_2$.

\noindent\textbf{Case 4} $P_i$ belongs to generation $G^{i_1}$, $P_j$ and $P_k$ belong to the same generation $G^{i_2}$, where $i_1<i_2$.
%\begin{figure}[!t]
%	\centering
%	\includegraphics[width=3.5in]{fig3.21.pdf}
%	\caption{The shares of $P_i$, $P_j$ and $P_k$ when $P_i$, $P_j$ and $P_k$ respectively belong to $G^{i_1}$, $G^{i_2}$ and $G^{i_3}$ with $i_1< i_2< i_3$.}
%	\label{fig3.21}
%\end{figure}

\noindent\textbf{Case 1} Let's consider the Case 1 first. $P_i$, $P_j$ and $P_k$ belong to different generations, we denote the corresponding generation respectively as $G^{i_1}$, $G^{i_2}$ and $G^{i_3}$ with $i_1< i_2< i_3$. All shares of $P_i$, $P_j$ and $P_k$ are shown as follows.

P1: ${sh}_{i_1}^{\infty}$, ${sh}_{i_2}^{\infty}$ and ${sh}_{i_3}^{\infty}$, are distributed by $\mathcal{S}^{\infty}$ among three different generations.

P2: There are $i_3-1$ forward shares ${sh}_{F}^{1}$, ${sh}_{F}^{2}$, $\cdots$, ${sh}_{F}^{i_3-1}$. The $i_1-1$ forward shares ${sh}_{F}^{1}$, ${sh}_{F}^{2}$, $\cdots$, ${sh}_{F}^{i_1-1}$ are the common shares of $P_i$, $P_j$ and $P_k$, the $i_2-i_1$ forward shares ${sh}_{F}^{i_1}$, $\cdots$, ${sh}_{F}^{i_2-1}$ are the common shares of $P_j$ and $P_k$, the $i_3-i_2$ forward shares ${sh}_{F}^{i_2}$, $\cdots$, ${sh}_{F}^{i_3-1}$ are provided separately by $P_k$. For $1\leq m \leq i_3-1$, each forward share ${sh}_{F}^{m}$ is one of shares constructed by the scheme $\mathcal{S}^{m}$.

P3: ${sh}_{h(i)}^{i_1}$, ${sh}_{h(j)}^{i_2}$, and ${sh}_{h(k)}^{i_3}$, the three shares are respectively constructed by the scheme $\mathcal{S}^{i_1}$, $\mathcal{S}^{i_2}$, and $\mathcal{S}^{i_3}$.

P4: There are $i_1+i_2+i_3-3$ backward shares ${SR}^{1}\oplus {sh}_{B}^{i_1}$, ${SR}^{2}\oplus {sh}_{B}^{i_1}$, $\cdots$, ${SR}^{i_1-1}\oplus {sh}_{B}^{i_1}$, ${SR}^{1}\oplus {sh}_{B}^{i_2}$, ${SR}^{2}\oplus {sh}_{B}^{i_2}$, $\cdots$, ${SR}^{i_2-1}\oplus {sh}_{B}^{i_2}$, ${SR}^{1}\oplus {sh}_{B}^{i_3}$, ${SR}^{2}\oplus {sh}_{B}^{i_3}$, $\cdots$, ${SR}^{i_3-1}\oplus {sh}_{B}^{i_3}$. The $i_1-1$ backward shares ${SR}^{1}\oplus {sh}_{B}^{i_1}$, ${SR}^{2}\oplus {sh}_{B}^{i_1}$, $\cdots$, ${SR}^{i_1-1}\oplus {sh}_{B}^{i_1}$ are provided  by $P_i$, the $i_2-1$ backward shares ${SR}^{1}\oplus {sh}_{B}^{i_2}$, ${SR}^{2}\oplus {sh}_{B}^{i_2}$, $\cdots$, ${SR}^{i_2-1}\oplus {sh}_{B}^{i_2}$ are provided by $P_j$, the $i_3-1$ backward shares ${SR}^{1}\oplus {sh}_{B}^{i_3}$, ${SR}^{2}\oplus {sh}_{B}^{i_3}$, $\cdots$, ${SR}^{i_3-1}\oplus {sh}_{B}^{i_3}$ are provided by $P_k$. For $m=i_1,i_2,i_3$, each forward share ${sh}_{B}^{m}$ is one of shares constructed by the scheme $\mathcal{S}^{m}$.

P5: Three random strings ${SR}^{i_1}$, ${SR}^{i_2}$ and ${SR}^{i_3}$.

Note that the shares of $P1$ consist of ${sh}_{i_1}^{\infty}, {sh}_{i_2}^{\infty}$ and ${sh}_{i_3}^{\infty}$, which are generated by the same evolving $3$-threshold scheme $\mathcal{S}^{\infty}$. Therefore, the three shares ${sh}_{i_1}^{\infty}, {sh}_{i_2}^{\infty}$ and ${sh}_{i_3}^{\infty}$ together can reconstruct the secret $s$ via $\mathcal{S}^{\infty}$. 

\noindent\textbf{Case 2} When $P_i$, $P_j$ and $P_k$ belong to the same generation, we denote the generation as $G^{i_1}$. According to the corresponding shares shown below.

P1: The share ${sh}_{i_1}^{\infty}$, is distributed by $\mathcal{S}^{\infty}$.

P2: There are $i_1-1$ forward shares ${sh}_{F}^{1}$, ${sh}_{F}^{2}$, $\cdots$, ${sh}_{F}^{i_1-1}$, which are the common shares of $P_i$, $P_j$ and $P_k$. For $1\leq m \leq i_1-1$, each forward share ${sh}_{F}^{m}$ is one of shares constructed by the scheme $\mathcal{S}^{m}$.

P3: ${sh}_{h(i)}^{i_1}$, ${sh}_{h(j)}^{i_1}$, and ${sh}_{h(k)}^{i_1}$, the three shares are constructed by the scheme $\mathcal{S}^{i_1}$.

P4: There are $i_1-1$ backward shares ${SR}^{1}\oplus {sh}_{B}^{i_1}$, ${SR}^{2}\oplus {sh}_{B}^{i_1}$, $\cdots$, ${SR}^{i_1-1}\oplus {sh}_{B}^{i_1}$, which are the common shares of $P_i$, $P_j$ and $P_k$. For $1\leq m \leq i_1-1$, each forward share ${sh}_{B}^{m}$ is one of shares constructed by the scheme $\mathcal{S}^{m}$.

P5: The random string ${SR}^{i_1}$.

Observing that the shares of $P3$ are composed of ${sh}_{h(i)}^{i_1}, {sh}_{h(j)}^{i_1}$ and ${sh}_{h(k)}^{i_1}$. They are generated by the same $(3,S(G^{i_1})+2)$-threshold scheme $\mathcal{S}^{i_1}$. Therefore, the three shares ${sh}_{h(i)}^{i_1}, {sh}_{h(j)}^{i_1}$ and ${sh}_{h(k)}^{i_1}$ can also reconstruct the secret $s$ via $\mathcal{S}^{i_1}$.
%\begin{figure}[!t]
%	\centering
%	\includegraphics[width=1.8in]{fig3.22.pdf}
%	\caption{The shares of $P_i$, $P_j$ and $P_k$ when $P_i$, $P_j$ and $P_k$ belong to the same generation $G^{i_1}$.}
%	\label{fig3.22}
%\end{figure}

\noindent\textbf{Case 3} When $P_i$ and $P_j$ belong to the same generation $G^{i_1}$, and $P_k$ belongs to another generation $G^{i_2}$ with $i_1<i_2$, the shares are shown as follows. 
 
P1: ${sh}_{i_1}^{\infty}$ and ${sh}_{i_2}^{\infty}$, are distributed by $\mathcal{S}^{\infty}$.

P2: There are $i_2-1$ forward shares ${sh}_{F}^{1}$, ${sh}_{F}^{2}$, $\cdots$, ${sh}_{F}^{i_2-1}$. The $i_1-1$ forward shares ${sh}_{F}^{1}$, ${sh}_{F}^{2}$, $\cdots$, ${sh}_{F}^{i_1-1}$ are the common shares of $P_i$, $P_j$ and $P_k$, the $i_2-i_1$ forward shares ${sh}_{F}^{i_1}$, $\cdots$, ${sh}_{F}^{i_2-1}$ are provided separately by $P_k$. For $1\leq m \leq i_2-1$, each forward share ${sh}_{F}^{m}$ is one of shares constructed by the scheme $\mathcal{S}^{m}$.

P3: ${sh}_{h(i)}^{i_1}$, ${sh}_{h(j)}^{i_1}$, and ${sh}_{h(k)}^{i_2}$, both ${sh}_{h(i)}^{i_1}$ and ${sh}_{h(j)}^{i_1}$ are constructed by the scheme $\mathcal{S}^{i_1}$, ${sh}_{h(k)}^{i_2}$ is constructed by the scheme $\mathcal{S}^{i_2}$.

P4: There are $i_1+i_2-2$ backward shares ${SR}^{1}\oplus {sh}_{B}^{i_1}$, ${SR}^{2}\oplus {sh}_{B}^{i_1}$, $\cdots$, ${SR}^{i_1-1}\oplus {sh}_{B}^{i_1}$, ${SR}^{1}\oplus {sh}_{B}^{i_2}$, ${SR}^{2}\oplus {sh}_{B}^{i_2}$, $\cdots$, ${SR}^{i_2-1}\oplus {sh}_{B}^{i_2}$. The $i_1-1$ backward shares ${SR}^{1}\oplus {sh}_{B}^{i_1}$, ${SR}^{2}\oplus {sh}_{B}^{i_1}$, $\cdots$, ${SR}^{i_1-1}\oplus {sh}_{B}^{i_1}$ are provided  by $P_i$ and $P_j$. The $i_2-1$ backward shares ${SR}^{1}\oplus {sh}_{B}^{i_2}$, ${SR}^{2}\oplus {sh}_{B}^{i_2}$, $\cdots$, ${SR}^{i_2-1}\oplus {sh}_{B}^{i_2}$ are provided by $P_k$. For $m=i_1,i_2$, each forward share ${sh}_{B}^{m}$ is one of shares constructed by the scheme $\mathcal{S}^{m}$.

P5: Two random strings ${SR}^{i_1}$ and ${SR}^{i_2}$.

Since $i_1<i_2$, for $P_k$, the $i_2-1$ forward shares include ${sh}_{F}^{i_1}$. On the other hand, for $P_i$ and $P_j$, the shares in $P3$ respectively are ${sh}_{h(i)}^{i_1}$ and ${sh}_{h(j)}^{i_1}$. The three shares ${sh}_{F}^{i_1}$, ${sh}_{h(i)}^{i_1}$ and ${sh}_{h(j)}^{i_1}$ are generated by the same $(3,S(G^{i_1})+2)$-threshold scheme $\mathcal{S}^{i_1}$, hence, $P_i$, $P_j$ and $P_k$ can reconstruct the secret $s$ by $\mathcal{S}^{i_1}$.

%\begin{figure}[!t]
%	\centering
%	\includegraphics[width=3.2in]{fig3.23.pdf}
%	\caption{The shares of $P_i$, $P_j$ and $P_k$ when $P_i$ and $P_j$ belong to $G^{i_1}$, and $P_k$ belong to $G^{i_2}$ with $i_1<i_2$.}
%	\label{fig3.23}
%\end{figure}

\noindent\textbf{Case 4} For the case that $P_i$ belongs to generation $G^{i_1}$, $P_j$ and $P_k$ belong to the same generation $G^{i_2}$ with $i_1<i_2$, the shares are shown below.
 
P1: ${sh}_{i_1}^{\infty}$ and ${sh}_{i_2}^{\infty}$, are distributed by $\mathcal{S}^{\infty}$.

P2: There are $i_2-1$ forward shares ${sh}_{F}^{1}$, ${sh}_{F}^{2}$, $\cdots$, ${sh}_{F}^{i_2-1}$. The $i_1-1$ forward shares ${sh}_{F}^{1}$, ${sh}_{F}^{2}$, $\cdots$, ${sh}_{F}^{i_1-1}$ are the common shares of $P_i$, $P_j$ and $P_k$, the $i_2-i_1$ forward shares ${sh}_{F}^{i_1}$, $\cdots$, ${sh}_{F}^{i_2-1}$ are the common shares of $P_j$ and $P_k$. For $1\leq m \leq i_2-1$, each forward share ${sh}_{F}^{m}$ is one of shares constructed by the scheme $\mathcal{S}^{m}$.

P3: ${sh}_{h(i)}^{i_1}$, ${sh}_{h(j)}^{i_2}$, and ${sh}_{h(k)}^{i_2}$, ${sh}_{h(i)}^{i_1}$ is constructed by the scheme $\mathcal{S}^{i_1}$, both ${sh}_{h(j)}^{i_2}$ and ${sh}_{h(k)}^{i_2}$ are constructed by the scheme $\mathcal{S}^{i_2}$. 

P4: There are $i_1+i_2-2$ backward shares ${SR}^{1}\oplus {sh}_{B}^{i_1}$, ${SR}^{2}\oplus {sh}_{B}^{i_1}$, $\cdots$, ${SR}^{i_1-1}\oplus {sh}_{B}^{i_1}$, ${SR}^{1}\oplus {sh}_{B}^{i_2}$, ${SR}^{2}\oplus {sh}_{B}^{i_2}$, $\cdots$, ${SR}^{i_2-1}\oplus {sh}_{B}^{i_2}$. The $i_1-1$ backward shares ${SR}^{1}\oplus {sh}_{B}^{i_1}$, ${SR}^{2}\oplus {sh}_{B}^{i_1}$, $\cdots$, ${SR}^{i_1-1}\oplus {sh}_{B}^{i_1}$ are provided by $P_i$. The $i_2-1$ backward shares ${SR}^{1}\oplus {sh}_{B}^{i_2}$, ${SR}^{2}\oplus {sh}_{B}^{i_2}$, $\cdots$, ${SR}^{i_2-1}\oplus {sh}_{B}^{i_2}$ are provided by $P_j$ and $P_k$. For $m=i_1,i_2$, each forward share ${sh}_{B}^{m}$ is one of shares constructed by the scheme $\mathcal{S}^{m}$.

P5: Two random strings ${SR}^{i_1}$ and ${SR}^{i_2}$.

For $P_i$, the share in $P5$ is ${SR}^{i_1}$. Since $P_j$ and $P_k$ belong to the same generation $G^{i_2}$, the $i_2-1$ backward shares in $P4$ are composed of  ${SR}^{1}\oplus {sh}_{B}^{i_2}, \cdots, {SR}^{i_2-1}\oplus {sh}_{B}^{i_2}$. As $i_1<i_2$, the $i_2-1$ backward shares include ${SR}^{i_1}\oplus {sh}_{B}^{i_2}$. Combining the share ${SR}^{i_1}$ provided by $P_i$ in $P5$, we can obtain ${sh}_{B}^{i_2}$ by calculating ${SR}^{i_1}\oplus {SR}^{i_1}\oplus {sh}_{B}^{i_2}$. On the other hand, for $P_j$ and $P_k$, the shares in $P3$ respectively are ${sh}_{h(j)}^{i_2}$ and ${sh}_{h(k)}^{i_2}$. The three shares ${sh}_{B}^{i_2}$, ${sh}_{h(j)}^{i_2}$ and ${sh}_{h(k)}^{i_2}$ are generated by the same  $(3,S(G^{i_2})+2)$-threshold scheme $\mathcal{S}^{i_2}$, hence $P_i$, $P_j$ and $P_k$ can reconstruct the secret $s$ by $\mathcal{S}^{i_2}$.
%\begin{figure}[!t]
%	\centering
%	\includegraphics[width=3.15in]{fig3.24.pdf}
%	\caption{The shares of $P_i$, $P_j$ and $P_k$ when $P_i$ belongs to $G^{i_1}$, $P_j$ and $P_k$ belong to $G^{i_2}$ with $i_1<i_2$.}
%	\label{fig3.24}
%\end{figure} 

Summing the above four cases, we show that any three participants can reconstruct the secret. 

\subsection{The Analysis of Secrecy}\label{sec3.3}
For any $C \in 2^{\mathcal{P}}\setminus\mathcal{A}$, we will prove that the secret $s$ is unable to be recovered by the shares of participants in $C$. $C$ is unqualified, then $|C|<3$. Since the cases of $|C|<3$ include the case of $|C|=2$, we only prove the case of $|C|=2$. Other cases can be similarly proved according to the case of $|C|=2$.

Denote the elements in $C$ as $P_i$ and $P_{j}$ with $i< j$. There exist two cases. (1) $P_i$ and $P_j$ belong to the same generation $G^{i_1}$. (2) $P_i$ and $P_j$ respectively belong to different generations $G^{i_1}$ and $G^{i_2}$ with $i_1<i_2$. 

When $P_i$ and $P_j$ belong to the same generation $G^{i_1}$, the all shares provided by $P_i$ and $P_j$ are below. 

P1: ${sh}_{i_1}^{\infty}$, is distributed by $\mathcal{S}^{\infty}$.

P2: There are $i_1-1$ forward shares ${sh}_{F}^{1}$, ${sh}_{F}^{2}$, $\cdots$, ${sh}_{F}^{i_1-1}$, which are the common shares of $P_i$ and $P_j$.

P3: ${sh}_{h(i)}^{i_1}$ and ${sh}_{h(j)}^{i_2}$, which are constructed by the scheme $\mathcal{S}^{i_1}$. 

P4: There are $i_1-1$ backward shares ${SR}^{1}\oplus {sh}_{B}^{i_1}$, ${SR}^{2}\oplus {sh}_{B}^{i_1}$, $\cdots$, ${SR}^{i_1-1}\oplus {sh}_{B}^{i_1}$, which are the common shares of $P_i$ and $P_j$.

P5: The random strings ${SR}^{i_1}$.

Observing these shares, we find that $P3$ is composed of $2$ shares ${sh}_{h(i)}^{i_1}$ and ${sh}_{h(j)}^{i_1}$, which are generated by the same $(3,S(G^{i_1})+2)$-threshold scheme $\mathcal{S}^{i_1}$. Therefore, if we want to construct the secret $s$ by using $\mathcal{S}^{i_1}$, we need to find another share that is also generated by $\mathcal{S}^{i_1}$. Note that only ${sh}_{F}^{i_1}$ or ${sh}_{B}^{i_1}$ can recover the secret together with ${sh}_{h(i)}^{i_2}$ and ${sh}_{h(i)}^{i_2}$. However, no matter how calculations are performed on these shares, it is impossible to construct ${sh}_{F}^{i_1}$ or ${sh}_{B}^{i_1}$. In this case, $P_i$ and $P_j$ can't obtain any information about the secret $s$.
%\begin{figure}[!t]
%	\centering
%	\includegraphics[width=2.0in]{fig3.31.pdf}
%	\caption{The shares of $P_i$ and $P_j$ when $P_i$ and $P_j$ belong to the same generation $G^{i_1}$.}
%	\label{fig3.31}
%\end{figure}

Consider the case that $P_i$ and $P_j$ respectively belong to different generations $G^{i_1}$ and $G^{i_2}$ with $i_1<i_2$. All shares provided by $P_i$ and $P_j$ are shown as follows. 

P1: ${sh}_{i_1}^{\infty}$ and ${sh}_{i_2}^{\infty}$, are distributed by $\mathcal{S}^{\infty}$.

P2: There are $i_2-1$ forward shares ${sh}_{F}^{1}$, ${sh}_{F}^{2}$, $\cdots$, ${sh}_{F}^{i_2-1}$. The $i_1-1$ forward shares ${sh}_{F}^{1}$, ${sh}_{F}^{2}$, $\cdots$, ${sh}_{F}^{i_1-1}$ are the common shares of $P_i$ and $P_j$, the $i_2-i_1$ forward shares ${sh}_{F}^{i_1}$, $\cdots$, ${sh}_{F}^{i_2-1}$ are provided by $P_j$. For $1\leq m \leq i_2-1$, each forward share ${sh}_{F}^{m}$ is one of shares constructed by the scheme $\mathcal{S}^{m}$.

P3: ${sh}_{h(i)}^{i_1}$ and ${sh}_{h(j)}^{i_2}$, which are constructed by the scheme $\mathcal{S}^{i_1}$ and $\mathcal{S}^{i_2}$, respectively. 

P4: There are $i_1+i_2-2$ backward shares ${SR}^{1}\oplus {sh}_{B}^{i_1}$, ${SR}^{2}\oplus {sh}_{B}^{i_1}$, $\cdots$, ${SR}^{i_1-1}\oplus {sh}_{B}^{i_1}$, ${SR}^{1}\oplus {sh}_{B}^{i_2}$, ${SR}^{2}\oplus {sh}_{B}^{i_2}$, $\cdots$, ${SR}^{i_2-1}\oplus {sh}_{B}^{i_2}$. The $i_1-1$ backward shares ${SR}^{1}\oplus {sh}_{B}^{i_1}$, ${SR}^{2}\oplus {sh}_{B}^{i_1}$, $\cdots$, ${SR}^{i_1-1}\oplus {sh}_{B}^{i_1}$ are provided by $P_i$. The $i_2-1$ backward shares ${SR}^{1}\oplus {sh}_{B}^{i_2}$, ${SR}^{2}\oplus {sh}_{B}^{i_2}$, $\cdots$, ${SR}^{i_2-1}\oplus {sh}_{B}^{i_2}$ are provided by $P_j$.

P5: Two random strings ${SR}^{i_1}$ and ${SR}^{i_2}$.

Notably, $P5$ contains the share ${SR}^{i_1}$, $P4$ of $P_j$ contains the share ${SR}^{i_1}\oplus {sh}_{B}^{i_2}$ since $1\leq i_1\leq i_2-1$. Thus, we can construct ${sh}_{B}^{i_2}$ by calculating ${SR}^{i_1}\oplus {SR}^{i_1}\oplus {sh}_{B}^{i_2}$. Since ${sh}_{B}^{i_2}$ is known, then each ${SR}^{k}$ satisfying $1\leq k\leq i_2-1$ can be obtained by calculating ${sh}_{B}^{i_2}\oplus {SR}^{k}\oplus {sh}_{B}^{i_2}$, where ${SR}^{k}\oplus {sh}_{B}^{i_2}$ is in $P4$. As $\{{SR}^{k}\}_{k=1}^{i_2-1}$ are known, selecting ${SR}^{1}$ from $\{{SR}^{k}\}_{k=1}^{i_2-1}$, then we can reconstruct ${sh}_{B}^{i_1}$ by calculating ${SR}^{1}\oplus {SR}^{1}\oplus {sh}_{B}^{i_1}$, where ${SR}^{1}\oplus {sh}_{B}^{i_1}$ is in $P4$. On the other hand, since $i_1\leq i_2-1$, $P2$ of $P_j$ contains the share ${sh}_{F}^{i_1}$. From $P3$ of $P_{i_1}$, we can obtain the share ${sh}_{h(i)}^{i_1}$. 
%\begin{figure}[!t]
%	\centering
%	\includegraphics[width=3.15in]{fig3.32.pdf}
%	\caption{The shares of $P_i$ and $P_j$ when $P_i$ and $P_j$ respectively belong to generations $G^{i_1}$ and $G^{i_2}$ with $i_1<i_2$.}
%	\label{fig3.32}
%\end{figure}

Now, we obtain three shares ${sh}_{B}^{i_1}$, ${sh}_{F}^{i_1}$ and ${sh}_{h(i)}^{i_1}$. They are generated by the same  $(3,S(G^{i_1})+2)$-threshold scheme $\mathcal{S}^{i_1}$, then $P_i$ and $P_j$ can reconstruct the secret $s$ by $\mathcal{S}^{i_1}$. Therefore, the proposed scheme is not secure.

However, the author of~\cite{DARCO2021149} got an incorrect conclusion when analyzing the security of this scheme, believing that the proposed scheme in \cite{DARCO2021149} has perfect security.

%%cheng!!!!

\section{The Revised Scheme with Perfect Secrecy Based on the Proposed Scheme in \cite{DARCO2021149}}\label{sec4}
As described in the previous section, the proposed scheme
is not secure. Considering the same conditions, we make slight modifications and propose a new evolving $3$-threshold secret sharing scheme with perfect secrecy. 

\subsection{Revised Scheme}\label{sub4.1}
Inspired by the construction of the scheme in \cite{DARCO2021149}, the scheme is also composed of an evolving $3$-threshold scheme $\mathcal{S}^{\infty}$ and a $3$-threshold scheme. Similarly, the evolving $3$-threshold scheme $\mathcal{S}^{\infty}$ is performed among different generations. Slightly difference is that the revised scheme uses a $(3,S(G^i)+1)$-threshold scheme $\mathcal{S}^{i}$ instead of $(3,S(G^i)+2)$-threshold scheme among the $i$-th generation. 

Denote $\mathcal P=\{P_1, P_2, \cdots, P_n, \cdots\}$ as the set of participants. When the participant joins the scheme, it is assigned to a corresponding generation. We recall the corresponding notations, which have already been given in Section~\ref{sub3.1}. For $i=1,2,\cdots$, the $i$-th generation denoted by $G^i$. Let $n_0=0$ and $n_i={2^{4^i}}$ for $i\geq 1$, $G^i$ begins when the $(n_{i-1}+1)$-th participant joins, and ends when the $n_i$-th participant joins. Therefore, the size of the $i$-th generation $G^i$ is $S(G^i)=n_i-n_{i-1}=2^{4^{i-1}}(2^4-1)$. The $t$-th participant belongs to the $g(t)$-th generation, where $g(t) = \lceil \log_4 \lg t\rceil$. Given the secret $s$, the evolving scheme $\mathcal{S}^{\infty}$ constructs many shares among different generations. For $i$-th generation with $S(G^i)$ participants, the $3$-threshold scheme $\mathcal{S}^{i}$ constructs $S(G^i)+1$ shares, denoted by ${sh}_{1}^{i}, {sh}_{2}^{i}, \cdots, {sh}_{S(G^i)}^{i}, {sh}_{F}^{i}$.

Similarly, for the $t$-th participant, we denote the corresponding index in the $g(t)$-th generation as $h(t)$, i.e. $1\leq h(t)\leq S(G^{g(t)})$. The corresponding shares of the $t$-th participant are composed of several pieces below.%, which are shown in Fig.~\ref{fig4.1}.

P1. The share ${sh}_{g(t)}^{\infty}$, which is distributed by $\mathcal{S}^{\infty}$ among different generations.

P2. There are $g(t)-1$ forward shares. For $j=1,2,\cdots,g(t)-1$, each forward share ${sh}_{F}^{j}$ is one of shares constructed by the scheme $\mathcal{S}^{j}$.

P3. One of these shares ${sh}_{1}^{g(t)}, {sh}_{2}^{g(t)}, \cdots, {sh}_{S(G^{g(t)})}^{g(t)}$, we denoted by ${sh}_{h(t)}^{g(t)}$ for $1\leq h(t)\leq S(G^{g(t)})$. These  shares and ${sh}_{F}^{g(t)}$ are constructed by the scheme $\mathcal{S}^{g(t)}$.

P4. There are $g(t)-1$ copies of the backward shares. For $1\leq j\leq g(t)-1$, each backward share masked with a different random string is described as ${SR}^{j}\oplus {sh}_{F}^{g(t)}$, where the share ${sh}_{F}^{g(t)}$ is constructed by the scheme $\mathcal{S}^{g(t)}$.

P5. A random string ${SR}^{g(t)}$.
%\begin{figure}[!t]
%	\centering
%	\includegraphics[width=2.75in]{fig4.1.pdf}
%	\caption{The whole shares of the $t$-th participant.}
%	\label{fig4.1}
%\end{figure}

\subsection{Proofs of Correctness and Secrecy}\label{sub4.2}
In this subsection, we will prove the correctness and secrecy of the proposed scheme.

\noindent\textbf{The proof of correctness} For any $A \in\mathcal{A}$, then $|A|\geq 3$. We need to show that the $\ell$-bit secret $s$ can be correctly reconstructed by the shares of the participants in $A$. We only prove the case of $|A|=3$. Without loss of generality, we denote three participants of $A$ as $P_i$, $P_j$ and $P_k$ with $i< j< k$. There also exist the following four cases.

\noindent\textbf{Case 1} $P_i$, $P_j$ and $P_k$ belong to different generations.

\noindent\textbf{Case 2} $P_i$, $P_j$ and $P_k$ belong to the same generation.

\noindent\textbf{Case 3} $P_i$ and $P_j$ belong to the same generation $G^{i_1}$, and $P_k$ belongs to another generation $G^{i_2}$, where $i_1<i_2$.

\noindent\textbf{Case 4} $P_i$ belongs to generation $G^{i_1}$, and $P_j$ and $P_k$ belong to the same generation $G^{i_2}$, where $i_1<i_2$.

The similar proof methods of Case 1, Case 2 and Case 3 have been mentioned in Section~\ref{sec3.3}. Therefore, we will no longer describe them again here.

Specifically, for the case that $P_i$ belongs to generation $G^{i_1}$, $P_j$ and $P_k$ belong to the same generation $G^{i_2}$ with $i_1<i_2$, we need consider it separately. All shares are described as follows. 

P1: ${sh}_{i_1}^{\infty}$ and ${sh}_{i_2}^{\infty}$, are distributed by $\mathcal{S}^{\infty}$.

P2: There are $i_2-1$ forward shares ${sh}_{F}^{1}$, ${sh}_{F}^{2}$, $\cdots$, ${sh}_{F}^{i_2-1}$. The $i_1-1$ forward shares ${sh}_{F}^{1}$, ${sh}_{F}^{2}$, $\cdots$, ${sh}_{F}^{i_1-1}$ are the common shares of $P_i$, $P_j$ and $P_k$, the $i_2-i_1$ forward shares ${sh}_{F}^{i_1}$, $\cdots$, ${sh}_{F}^{i_2-1}$ are the common shares of $P_j$ and $P_k$. For $1\leq m \leq i_2-1$, each forward share ${sh}_{F}^{m}$ is one of shares constructed by the scheme $\mathcal{S}^{m}$.

P3: ${sh}_{h(i)}^{i_1}$, ${sh}_{h(j)}^{i_2}$, and ${sh}_{h(k)}^{i_2}$, ${sh}_{h(i)}^{i_1}$ is constructed by the scheme $\mathcal{S}^{i_1}$, both ${sh}_{h(j)}^{i_2}$ and ${sh}_{h(k)}^{i_2}$ are constructed by the scheme $\mathcal{S}^{i_2}$. 

P4: There are $i_1+i_2-2$ backward shares ${SR}^{1}\oplus {sh}_{F}^{i_1}$, ${SR}^{2}\oplus {sh}_{F}^{i_1}$, $\cdots$, ${SR}^{i_1-1}\oplus {sh}_{F}^{i_1}$, ${SR}^{1}\oplus {sh}_{F}^{i_2}$, ${SR}^{2}\oplus {sh}_{F}^{i_2}$, $\cdots$, ${SR}^{i_2-1}\oplus {sh}_{F}^{i_2}$. The $i_1-1$ backward shares ${SR}^{1}\oplus {sh}_{F}^{i_1}$, ${SR}^{2}\oplus {sh}_{F}^{i_1}$, $\cdots$, ${SR}^{i_1-1}\oplus {sh}_{F}^{i_1}$ are provided by $P_i$. The $i_2-1$ backward shares ${SR}^{1}\oplus {sh}_{F}^{i_2}$, ${SR}^{2}\oplus {sh}_{F}^{i_2}$, $\cdots$, ${SR}^{i_2-1}\oplus {sh}_{F}^{i_2}$ are provided by $P_j$ and $P_k$. For $m=i_1,i_2$, each forward share ${sh}_{F}^{m}$ is one of shares constructed by the scheme $\mathcal{S}^{m}$.

P5: Two random strings ${SR}^{i_1}$ and ${SR}^{i_2}$.

The share in $P5$ of $P_i$ is the binary string ${SR}^{i_1}$. Since $P_j$ and $P_k$ belong to the same generation $G^{i_2}$, the $i_2-1$ backward shares in $P4$ for $P_j$ and $P_k$ are the same, which are ${SR}^{1}\oplus {sh}_{F}^{i_2}, \cdots, {SR}^{i_2-1}\oplus {sh}_{F}^{i_2}$. As $i_1<i_2$, the $i_2-1$ backward shares include ${SR}^{i_1}\oplus {sh}_{F}^{i_2}$. Combining the share ${SR}^{i_1}$ provided by $P_i$, we can obtain ${sh}_{F}^{i_2}$ by calculating ${SR}^{i_1}\oplus {SR^{i_1}}\oplus {sh}_{F}^{i_2}$. On the other hand, for $P_j$ and $P_k$, the shares in $P3$ respectively are ${sh}_{h(j)}^{i_2}$ and ${sh}_{h(k)}^{i_2}$. The three shares ${sh}_{F}^{i_2}$, ${sh}_{h(j)}^{i_2}$ and ${sh}_{h(k)}^{i_2}$ are generated by the $(3,S(G^{i_2})+1)$-threshold scheme $\mathcal{S}^{i_2}$, therefore $P_i$, $P_j$ and $P_k$ can reconstruct the secret $s$ by $\mathcal{S}^{i_2}$.
%\begin{figure}[!t]
%	\centering
%	\includegraphics[width=3.15in]{fig4.21.pdf}
%	\caption{The shares of $P_i$, $P_j$ and $P_k$ when $P_i$ belongs to $G^{i_1}$, $P_j$ and $P_k$ belong to $G^{i_2}$ with $i_1<i_2$.}
%	\label{fig4.21}
%\end{figure}

\noindent\textbf{The proof of secrecy} For any $C \in 2^{\mathcal{P}}\setminus\mathcal{A}$, we will prove that the secret $s$ is unable to be recovered by the shares of participants in $C$. $C$ is unqualified, then $|C|<3$. We only prove the case of $|C|=2$. Denote the elements in $C$ as $P_i$ and $P_{j}$ with $i< j$. There exist two cases. (1) $P_i$ and $P_j$ belong to the same generation. (2) $P_i$ and $P_j$ respectively belong to different generations. 

%\begin{figure}[!t]
%	\centering
%	\includegraphics[width=2.95in]{fig4.22.pdf}
%	\caption{The shares of $P_i$ and $P_j$ for $P_i$ and $P_j$ respectively belong to generations $G^{i_1}$ and $G^{i_2}$ with $i_1<i_2$.}
%	\label{fig4.22}
%\end{figure}

For the first case, the similar proof have been shown in Section~\ref{sec3.3}. Therefore, we will no longer prove it again here. When $P_i$ and $P_j$ respectively belong to different generations $G^{i_1}$ and $G^{i_2}$ with $i_1<i_2$. All shares provided by $P_i$ and $P_j$ are shown as follows.

P1: ${sh}_{i_1}^{\infty}$ and ${sh}_{i_2}^{\infty}$, are distributed by $\mathcal{S}^{\infty}$.

P2: There are $i_2-1$ forward shares ${sh}_{F}^{1}$, ${sh}_{F}^{2}$, $\cdots$, ${sh}_{F}^{i_2-1}$. The $i_1-1$ forward shares ${sh}_{F}^{1}$, ${sh}_{F}^{2}$, $\cdots$, ${sh}_{F}^{i_1-1}$ are the common shares of $P_i$ and $P_j$, the $i_2-i_1$ forward shares ${sh}_{F}^{i_1}$, $\cdots$, ${sh}_{F}^{i_2-1}$ are provided by $P_j$. For $1\leq m \leq i_2-1$, each forward share ${sh}_{F}^{m}$ is one of shares constructed by the scheme $\mathcal{S}^{m}$.

P3: ${sh}_{h(i)}^{i_1}$ and ${sh}_{h(j)}^{i_2}$, which are constructed by the scheme $\mathcal{S}^{i_1}$ and $\mathcal{S}^{i_2}$, respectively. 

P4: There are $i_1+i_2-2$ backward shares ${SR}^{1}\oplus {sh}_{F}^{i_1}$, ${SR}^{2}\oplus {sh}_{F}^{i_1}$, $\cdots$, ${SR}^{i_1-1}\oplus {sh}_{F}^{i_1}$, ${SR}^{1}\oplus {sh}_{F}^{i_2}$, ${SR}^{2}\oplus {sh}_{F}^{i_2}$, $\cdots$, ${SR}^{i_2-1}\oplus {sh}_{F}^{i_2}$. The $i_1-1$ backward shares ${SR}^{1}\oplus {sh}_{F}^{i_1}$, ${SR}^{2}\oplus {sh}_{F}^{i_1}$, $\cdots$, ${SR}^{i_1-1}\oplus {sh}_{F}^{i_1}$ are provided by $P_i$. The $i_2-1$ backward shares ${SR}^{1}\oplus {sh}_{B}^{i_2}$, ${SR}^{2}\oplus {sh}_{F}^{i_2}$, $\cdots$, ${SR}^{i_2-1}\oplus {sh}_{F}^{i_2}$ are provided by $P_j$.

P5: Two random strings ${SR}^{i_1}$ and ${SR}^{i_2}$.

Notably, there only exist two shares ${sh}_{i_1}^{\infty}$ and ${sh}_{i_2}^{\infty}$ in $P1$, which is not enough to reconstruct the secret $s$ by $\mathcal{S}^{\infty}$. Consider the scheme $\mathcal{S}^{i_1}$, we can only obtain two shares ${sh}_{h(i)}^{i_1}$ provided by $P3$ and ${sh}_{F}^{i_1}$ provided by $P2$. However, except ${sh}_{F}^{i_1}$ and ${sh}_{h(i)}^{i_1}$, we can't find out another share generated by $\mathcal{S}^{i_1}$. Therefore, the two shares ${sh}_{F}^{i_1}$ and ${sh}_{h(i)}^{i_1}$ are not enough to obtain any information about $s$. Consider that another scheme $\mathcal{S}^{i_2}$, $P5$ can provide a share ${SR}^{i_1}$ and $P4$ can provide the share ${SR}^{i_1}\oplus {sh}_{F}^{i_2}$ since $1\leq i_1\leq i_2-1$. Thus, we can calculate ${sh}_{F}^{i_2}$ by calculating ${SR}^{i_1}\oplus {SR}^{i_1}\oplus {sh}_{F}^{i_2}$. Combining the share ${sh}_{h(j)}^{i_2}$ provided by $P3$, we get two shares ${sh}_{F}^{i_2}$ and ${sh}_{h(j)}^{i_2}$, which are generated by $\mathcal{S}^{i_2}$. However, except ${sh}_{F}^{i_2}$ and ${sh}_{h(j)}^{i_2}$, we can't find out another share generated by $\mathcal{S}^{i_2}$. Therefore, the two shares ${sh}_{F}^{i_2}$ and ${sh}_{h(j)}^{i_2}$ are also not enough to obtain any information about $s$. 

Summing the above two cases, arbitrary two participants can't obtain any information about the secret $s$. Therefore, the revised scheme has perfect secrecy.

\subsection{The Analysis of Share Size}
In this part, we analyze the share size of each participant. For given the secret $s$, $Z^{(s)}_{P_t}$ represents all shares of the $t$-th participant and $B(a)$ represents the corresponding size of the share $a$. The shares of the $t$-th participant are composed of five parts $P1$, $P2$, $P3$, $P4$ and $P5$, where $g(t)\geq 2$. Therefore, $B(Z^{(s)}_{P_t})$ is equal to the sum of all shares’ size in $P1$, $P2$, $P3$, $P4$ and $P5$. For the convenience of description, denote the share size of all shares in $Pi$ as $B(Pi)$, where $1\leq i\leq 5$. Then, we have
\begin{equation}\label{key431}
	B(Z^{(s)}_{P_t})=B(P1)+B(P2)+B(P3)+B(P4)+B(P5).
\end{equation}
In (\ref{key431}), $B(P1)$ is equal to $B({sh}_{g(t)}^{\infty})$. The share ${sh}_{g(t)}^{\infty}$ is constructed by the scheme $\mathcal{S}^{\infty}$ with the share size as $O(g^2(t))$, which is mentioned in~\cite{DARCO2021149}. Since $g(t)=\lceil \log_4 \lg t\rceil$, then
\begin{equation}\label{key432}
	B(P1)=B({sh}_{g(t)}^{\infty})=O(g^2(t))=O({\lceil \log_4 \lg t\rceil}^2).
\end{equation}

Next, we will calculate $B(P2)$, the value of $B(P2)$ is equal to $\sum_{j=1}^{g(t)-1}B({sh}_{F}^{j})$. Each share ${sh}_{F}^{j}$ is constructed by the scheme $\mathcal{S}^{j}$ for any $j$ with $1\leq j\leq g(t)-1$. Besides, we find that $B(P4)=\sum_{j=1}^{g(t)-1}B({SR}^{j}\oplus {sh}_{F}^{g(t)})$, each share ${SR}^{j}\oplus {sh}_{F}^{g(t)}$ is that the random string ${SR}^{j}$ masked with the fixed share ${sh}_{F}^{g(t)}$ generated by the scheme $\mathcal{S}^{g(t)}$. Therefore, the share size of each ${SR}^{j}$ is equal to $B({sh}_{F}^{g(t)})$. For each participant of the $j$-th generation with $1\leq j\leq g(t)-1$, the shares in $P4$ contain ${SR}^{1}\oplus {sh}_{F}^{j}$ for the random string ${SR}^{1}$, where ${sh}_{F}^{j}$ is generated by the scheme $\mathcal{S}^{j}$. Once ${SR}^{1}$ is selected, it is fixed, therefore, the size of each ${sh}_{F}^{j}$ is the same, where $2\leq j\leq g(t)-1$. Since the share size of each ${SR}^{j}$ is equal to $B({sh}_{F}^{g(t)})$, combining the above fact, we assert that the size of any share ${sh}_{F}^{i}$ and ${SR}^{k}$ is equal for $i,k\geq 1$. Therefore, we have
\begin{equation}\label{key433}
	\begin{aligned}
		&B(P2)+B(P4)+B(P5)\\
		=&\sum_{j=1}^{g(t)-1}B({sh}_{F}^{j})
		+\sum_{j=1}^{g(t)-1}B({SR}^{j}\oplus {sh}_{F}^{g(t)})+B({SR}^{g(t)})\\
		=&\big(g(t)-1\big)\cdot B({sh}_{F}^{1})+\big(g(t)-1\big)\cdot B({sh}_{F}^{1})+B({SR}^{g(t)})\\
		=&\big(2g(t)-1\big)B({sh}_{F}^{1}).
	\end{aligned}
\end{equation}

The authors of~\cite{DARCO2021149} have designed a $(3,S(G^i)+2)$-threshold scheme $\mathcal{S}^{i}$ based on the Chinese reminder theorem, and achieved the goal that the sizes of any ${sh}_{F}^{j}$ and ${sh}_{B}^{k}$ are $\lg p$ with $j,k\geq1$ and $p$ is a prime. By making slight changes, we can propose a $(3,S(G^i)+1)$-threshold scheme $\mathcal{S}^{i}$ based on the Chinese reminder theorem, with the size of ${sh}_{F}^{j}$ as $B({sh}_{F}^{j})=\lg p$, where $i\geq1$. Taking the result into \eqref{key433}, then we have 
\begin{equation}\label{key4331}
	\begin{aligned}
	&B(P2)+B(P4)+B(P5)\\
	=&\big(2g(t)-1\big)B({sh}_{F}^{1})\\
    =&\big(2g(t)-1\big)\cdot \lg p\\
    =&\lg p (2\lceil \log_4 \lg t\rceil-1).
    \end{aligned}
\end{equation}

As for $B(P3)$, it is equal to $B({sh}_{h(t)}^{g(t)})$. The scheme $\mathcal{S}^{g(t)}$ constructs the share ${sh}_{h(t)}^{g(t)}$ with the share size approximately as $\lg t$, which is described in \cite{DARCO2021149}. Combining the values of $B(P1)$, $B(P2)+B(P4)+B(P5)$ and $B(P3)$, we obtain
\begin{equation}\label{key434}
   \begin{aligned}
  B(Z^{(s)}_{P_t})=&B(P1)+B(P2)+B(P3)+B(P4)+B(P5)\\
  =&B(P1)+B(P2)+B(P4)+B(P5)+B(P3)\\
    =&O({\lceil \log_4 \lg t\rceil}^2)+\lg p(2\lceil \log_4 \lg t\rceil-1)+\lg t\\
    =&\lg t+O({\lceil \log_4 \lg t\rceil}^2)+\lg p(2\lceil \log_4 \lg t\rceil-1).	
   \end{aligned}	
\end{equation}
From (\ref{key434}), the most important term of $B(Z^{(s)}_{P_t})$ is $\lg t$. However, the most important term of $B(Z^{(s)}_{P_t})$ is $2\lg t$ in Theorem~\ref{thm1}. By comparing, the share size of the proposed scheme is smaller than the result of the scheme in~\cite{komargodski2016share,komargodski2017share}.

\section{A New Construction of Evolving $3$-threshold Secret Sharing Scheme}\label{sec5}
As described in Section~\ref{sub4.1}, we have given the model of evolving $3$-threshold secret sharing scheme with perfect security. The scheme is composed of a normal evolving $3$-threshold scheme $\mathcal{S}^{\infty}$ among different generations and a $(3,S(G^i)+1)$-threshold scheme $\mathcal{S}^{i}$ among the $i$-th generation, where $S(G^i)$ denote the number of participants for the $i$-th generation. In the following, we propose a new method to construct $(3,S(G^i)+1)$-threshold scheme $\mathcal{S}^{i}$.

\subsection{The New Design of Conventional $3$-threshold Secret Sharing Scheme over the Finite Field $F_{2^{\ell m}}$}\label{sec5.1}
In this subsection, based on Shamir's scheme~\cite{shamir1979share}, we propose a new construction of $(3,2^{\ell m-1}+1)$ secret sharing scheme with over the finite field $F_{2^{\ell m}}$ for an $\ell$-bit secret $s$, where $m,\ell \in\mathbb{N}^+$.

\subsubsection{Proposed Scheme}\label{sub5.1}
For any $\ell, m\in\mathbb{N}^+$, we describe two finite fields $F_{2^\ell}$ and $F_{2^{\ell m}}$, which are used in the scheme. We first consider the finite field $F_{2^\ell}$. It is isomorphic to a polynomial quotient ring $F_2[x]/g(x)$, where $g(x)$ is an monic irreducible polynomial in $F_2[x]$ of degree $\ell$. $F_{2^\ell}$ contains $2^\ell$ elements. For any integer $i$ with $0\leq i\leq 2^\ell-1$, it's binary representation $c_i$ is denoted as $c_i=(c_{i,0},c_{i,1},\cdots,c_{i,{\lfloor\lg i\rfloor}})$, where $\lfloor\lg i\rfloor+1$ represents the length of $c_i$ with $\lfloor\lg i\rfloor+1\leq \ell$. Let $\alpha_i$ denote the polynomial form of $c_i$ in $F_2[x]/g(x)$, which is defined as
\begin{equation*}
	\alpha_i=\sum_{k=0}^{\lfloor\lg i\rfloor}c_{i,k}x^k,
\end{equation*} 
where $c_{i,k}\in F_2$. The finite field $F_{2^\ell}$ is composed of such $\alpha_i$ for $0\leq i\leq 2^\ell-1$.

For any $m\in\mathbb{N}^+$, we construct the extension field $F_{2^{\ell m}}$ over $F_{2^\ell}$. It is isomorphic to a polynomial quotient ring $F_{2^\ell}[y]/g_1(y)$, where $g_1(y)$ is an monic irreducible polynomial in $F_{2^\ell}[y]$ of degree $m$. $F_{2^{\ell m}}$ contains $2^{\ell m}$ elements. For any integer $j$ with $0\leq j\leq 2^{\ell m}-1$, it's binary representation $c_j$ is denoted as $c_j=(c_{j,0},c_{j,1},\cdots,c_{j,{\lfloor\lg j\rfloor}})$, where $\lfloor\lg j\rfloor+1$ represents the length of $c_j$ with $\lfloor\lg j\rfloor+1\leq \ell m$. Let $\beta_j$ denote the polynomial form of $c_j$ in $F_{2^\ell}[y]/g_1(y)$, which is defined as
\begin{equation*}
	\beta_j=\sum_{k=0}^{m-1}c_{j,k}y^k,
\end{equation*} 
where $c_{j,k}\in F_{2^\ell}$ is denoted as the polynomial form of the part binary string $(c_{j,k\ell},c_{j,k\ell+1},\cdots,c_{j,(k+1)\ell-1})$ in $c_j$.
The finite field $F_{2^{\ell m}}$ is composed of such $\beta_j$ for $0\leq j\leq 2^{\ell m}-1$. Notably, we may mix the two representions $F_{2^\ell}$ ($F_{2^{\ell m}}$) and $F_2[x]/g(x)$ ($F_{2^\ell m}[y]/g_1(y)$) about the finite field in following content.

Next, we show the proposed scheme. In the finite field $F_{2^\ell}[y]/g_1(y)$, given an $\ell$-bit binary secret $s$, for the $(i+1)$-th participant $P_{i+1}$ with $0\leq i\leq 2^{\ell m-1}-1$, the algorithm $\mathcal{E}$ constructs the corresponding share $Z_i$ as 
\begin{equation}\label{key301}
	Z_i=F(2i)=F(\beta_{2i})=a_0+a_1\beta_{2i}+a_2(\beta_{2i})^2,
\end{equation}
where $a_0, a_1$ are randomly chosen in $F_{2^\ell}[y]/g_1(y)$, $a_2\in F_2[x]/g(x)$ is defined as
\begin{equation}\label{key302}
	a_2=a_1+s.
\end{equation}
Notably, $s$ in (\ref{key302}) uses the polynomial form, which would be the default form throughout this paper. Actually, though $s \in F_2[x]/g(x)$, it is also the element of the extension field $F_{2^\ell}[y]/g_1(y)$. Hence the value of $a_1+s$ is in $F_{2^\ell}[y]/g_1(y)$, which can be represented as 
\begin{equation*}
	a_1+s=\sum_{k=0}^{m-1}\alpha_{a,k} y^k,
\end{equation*}
where $\alpha_{a,k} \in F_2[x]/g(x)$. To make $a_2\in F_2[x]/g(x)$, the value of $a_2$ can be seen as a natural mapping by $a_1+s$ from $F_{2^\ell}[y]/g_1(y)$ to $F_2[x]/g(x)$, i.e. $a_2=\alpha_{a,0}$.

In the proposed scheme, the share space is composed of  $\{F(2i)\}_{i=0}^{2^{\ell m-1}-1}\cup\{a_2\}$. Consider the finite field $F_{2^\ell}[y]/g_1(y)$ with $2^{\ell m}$ elements, we select $2^{\ell m-1}$ even numbers and make the corresponding function values as the shares, instead of selecting odd numbers. This reason will be explained later.

\noindent\textbf{Discussion} Now, we explain the reasons why the share space is composed of $\{F(2i)\}_{i=0}^{2^{\ell m-1}-1}\cup\{a_2\}$ instead of $\{F(j)\}_{j=0}^{2^{\ell m}-1}\cup\{a_2\}$. 

Supposing the share space is composed of $\{F(j)\}_{j=0}^{2^{\ell m}-1}\cup\{a_2\}$. For any $s$, we have
\begin{equation*}
	Z_i=a_0+a_1(\beta_{i})+a_2(\beta_{i})^2.	
\end{equation*}
where $a_0, a_1$ are randomly chosen in $F_{2^\ell}[y]/g_1(y)$, $a_2\in F_2[x]/g(x)$ is defined as
\begin{equation*}
	a_2=a_1+s.
\end{equation*}

For the convenience of discussion, let ${\mathcal{Q}}=\{F(2i)\}_{i=0}^{2^{\ell m-1}-1}$. Selecting any two participants $P_{i+1}$ and $P_{j+1}$, when the corresponding shares $Z_i$ and $Z_j$ belong to ${\mathcal{Q}}$, there may exist the case $i+j=1 \pmod 2$. 

For the given $(\beta_{i}, Z_i)$ and $(\beta_{j}, Z_j)$, we have
\begin{equation}\label{key3211}
	\begin{aligned}
		\left
		\{
		\begin{array}{c}
			Z_i=F(\beta_{i})=a_0+a_1\beta_{i}+a_2(\beta_{i})^2,\\
			Z_j=F(\beta_{j})=a_0+a_1\beta_{j}+a_2(\beta_{j})^2,\\
		\end{array} \right.
	\end{aligned}
\end{equation}
we subtract the first equation from the second equation in \eqref{key3211} and get 
\begin{equation}\label{key3231}
	Z_j-Z_i=a_1(\beta_{j}-\beta_{i})+a_2(\beta_{j}-\beta_{i})(\beta_{j}+\beta_{i}).
\end{equation}
In the finite field $F_{2^\ell}[y]/g_1(y)$, dividing both sides of the equation by $\beta_{j}-\beta_{i}$, then we have 
\begin{equation}\label{key3241}
	\frac{Z_j-Z_i}{\beta_{j}-\beta_{i}}=a_1+a_2(\beta_{j}+\beta_{i}).
\end{equation}
Let $a_1=a_{\ell}+a_{h}y$, then $a_2=a_{\ell}+s$, where $a_{\ell}\in F_{2^\ell}$ and $a_{h}\in F_{2^{\ell(m-1)}}$. Substituting $a_{1}=a_{\ell}+a_{h}y$ and $a_{2}=a_{\ell}+s$ into the above equation, we can get 
\begin{align}\label{key3251}
	\frac{Z_j-Z_i}{\beta_{j}-\beta_{i}}=&a_{1}+a_{2}(\beta_{j}+\beta_{i})\nonumber\\
	=&a_{\ell}+a_{h}y+(a_{\ell}+s)(\beta_{j}+\beta_{i})\nonumber\\
	=&a_{\ell}(1+\beta_{j}+\beta_{i})+a_{h}y+s(\beta_{j}+\beta_{i}).
\end{align}
From \eqref{key3251}, we can further derive
\begin{equation}\label{key3252}
	\frac{Z_j-Z_i}{\beta_{j}-\beta_{i}}\equiv a_{\ell}(1+\beta_{j}+\beta_{i})+s(\beta_{j}+\beta_{i}) \pmod y.
\end{equation}
Since $i+j=1 \pmod 2$, then the first bit of $\beta_{j}+\beta_{i}$ is $1$. Therefore, we can calculate that the first bit of $s$ is equal to the first bit of $\frac{Z_j-Z_i}{\beta_{j}-\beta_{i}}$. Thus, we can obtain some information about the secret, then the setting of that the share space composed by $\{F(j)\}_{j=0}^{2^{\ell m}-1}\cup\{a_2\}$ is unreasonable. 

\subsubsection{Proofs of Correctness and Secrecy}\label{sub5.2}
In this subsection, we will prove the correctness and secrecy of the proposed scheme. As described in the proposed scheme, the share space is composed of $\{F(2i)\}_{i=0}^{2^{\ell m-1}-1}\cup\{a_2\}$. For the convenience of writing, let ${\mathcal{Z}}=\{F(2i)\}_{i=0}^{2^{\ell m-1}-1}$.

\noindent\textbf{Correctness} For any $A \in\mathcal{A}$, then $|A|\geq 3$. We need to show that the $\ell$-bit secret $s$ can be correctly reconstructed by the shares of the participants in $A$. Similarly, we only prove the case of $|A|=3$.  

Without loss of generality, we denote the three participants as $P_{i+1}$, $P_{j+1}$ and $P_{k+1}$ in $A$ with $0\leq i<j<k\leq 2^{\ell m}-1$. Considering the corresponding secret shares $Z_i$, $Z_j$ and $Z_k$, there exist two following cases.

\noindent\textbf{Case 1} $Z_i$, $Z_j$ and $Z_k$ are in ${\mathcal{Z}}$.

\noindent\textbf{Case 2} Two shares are in ${\mathcal{Z}}$ and the rest one is equal to $a_2$.

\noindent\textbf{The proof of Case 1} Consider the quadratic function $F(w)=a_0+a_1w+a_2w^2$ in the finite field $F_{2^{\ell m}}$. Let $\beta_{2i}$, $\beta_{2j}$ and $\beta_{2k}$ respectively represent the corresponding polynomial forms of integers $2i$, $2j$ and $2k$. For given three pairs of values $(\beta_{2i}, Z_i)$, $(\beta_{2j}, Z_j)$ and $(\beta_{2k}, Z_k)$, the following equations
\begin{equation}\label{key303}
	\begin{aligned}
		\left
		\{
		\begin{array}{c}
			Z_i=F(\beta_{2i})=a_0+a_1\beta_{2i}+a_2(\beta_{2i})^2,\\
			Z_j=F(\beta_{2j})=a_0+a_1\beta_{2j}+a_2(\beta_{2j})^2,\\
			Z_k=F(\beta_{2k})=a_0+a_1\beta_{2k}+a_2(\beta_{2k})^2,\\
		\end{array} \right.
	\end{aligned}
\end{equation}
hold. 

According to the Lagrange interpolation formula, we can obtain
\begin{equation}\label{key304}
	\begin{aligned}
		F(w)=&Z_i\frac{(w-\beta_{2j})(w-\beta_{2k})}{(\beta_{2i}-\beta_{2j})(\beta_{2i}-\beta_{2k})}\\
		+&Z_j\frac{(w-\beta_{2i})(w-\beta_{2k})}{(\beta_{2j}-\beta_{2i})(\beta_{2j}-\beta_{2k})}\\
		+&Z_k\frac{(w-\beta_{2i})(w-\beta_{2j})}{(\beta_{2k}-\beta_{2i})(\beta_{2k}-\beta_{2j})}.	
	\end{aligned}
\end{equation}
By expanding the above quadratic function and merging similar terms, we can obtain the solutions of $a_0$, $a_1$ and $a_2$.

Since the secret $s$ is equal to the value of $a_1+a_2$ over the finite field $F_2[x]/g(x)$, substituting the results of $a_1$ and $a_2$ into $a_1+a_2$, then we can construct the secret $s$ by computing the value of $a_1+a_2$ over $F_2[x]/g(x)$ as follows
\begin{align}\label{key306}	
	s=&a_2+a_1\nonumber\\
	=&\frac{(\beta_{2j}+\beta_{2k}-1)(\beta_{2k}-\beta_{2j})\cdot Z_i}{(\beta_{2i}-\beta_{2j})(\beta_{2i}-\beta_{2k})(\beta_{2j}-\beta_{2k})}\nonumber\\
	+&\frac{(\beta_{2i}+\beta_{2k}-1)(\beta_{2i}-\beta_{2k})\cdot Z_j}{(\beta_{2i}-\beta_{2j})(\beta_{2i}-\beta_{2k})(\beta_{2j}-\beta_{2k})}\nonumber\\
	+&\frac{(\beta_{2i}+\beta_{2j}-1)(\beta_{2j}-\beta_{2i})\cdot Z_k}{(\beta_{2i}-\beta_{2j})(\beta_{2i}-\beta_{2k})(\beta_{2j}-\beta_{2k})}.
\end{align}
\noindent\textbf{The proof of Case 2} Without loss of generality, suppose $Z_i, Z_j \in {\mathcal{Z}}$ and $Z_k=a_2$. Consider a quadratic function $F(w)=a_0+a_1w+a_2w^2$ in the finite field $F_{2^{\ell m}}$. For given two pairs of values $(\beta_{2i}, Z_i)$ and $(\beta_{2j}, Z_j)$, the following equations
\begin{equation}\label{key307}
	\begin{aligned}
		\left
		\{
		\begin{array}{c}
			Z_i=F(\beta_{2i})=a_0+a_1\beta_{2i}+a_2(\beta_{2i})^2,\\
			Z_j=F(\beta_{2j})=a_0+a_1\beta_{2j}+a_2(\beta_{2j})^2,\\
		\end{array} \right.
	\end{aligned}
\end{equation}
hold. 

Since $a_2$ is known, let $\bar{F}(w)=F(w)-a_2w^2$, then the above equations become
\begin{equation}\label{key308}
	\begin{aligned}
		\left
		\{
		\begin{array}{c}
			\bar{F}(\beta_{2i})=Z_i-a_2(\beta_{2i})^2=a_0+a_1\beta_{2i},\\
			\bar{F}(\beta_{2j})=Z_j-a_2(\beta_{2j})^2=a_0+a_1\beta_{2j}.\\
		\end{array} \right.
	\end{aligned}
\end{equation}
For known $(\beta_{2i}, Z_i-a_2(\beta_{2i})^2)$ and $(\beta_{2j},Z_j-a_2(\beta_{2j})^2)$,
using the Lagrange interpolation formula, we can get the expression of $\bar{F}(w)$, i.e.
\begin{equation}\label{key309}
	\begin{aligned}
		\bar{F}(w)=&\Big(Z_i-a_2 (\beta_{2i})^2\Big)\frac{w-\beta_{2j}}{\beta_{2i}-\beta_{2j}}\\
		+&\Big(Z^{(s)}_{j}-a_2(\beta_{2j})^2\Big)\frac{w-\beta_{2i}}{\beta_{2j}-\beta_{2i}}.	
	\end{aligned}
\end{equation}
By expanding the above function and merging similar terms, we have
\begin{equation}\label{key310}
	\begin{aligned}
		\left
		\{
		\begin{array}{l}
			a_0=\frac{\beta_{2j}\big(a_2 (\beta_{2i})^2-Z_i\big)+\beta_{2i}\big(Z_j-a_2 (\beta_{2j})^2\big)}{\beta_{2i}-\beta_{2j}},\\
			a_1=\frac{\big(Z_{i}-a_2(\beta_{2i})^2\big)-\big(Z_{j}-a_2(\beta_{2j})^2\big)}{\beta_{2i}-\beta_{2j}}.\\
		\end{array} \right.
	\end{aligned}
\end{equation}
Since the secret $s$ is equal to the value of $a_1+a_2$ over the finite field $F_2[x]/g(x)$, substituting the results of $a_1$ and $a_2$ into $a_1+a_2$, then we can construct the secret $s$ by computing the value of $a_1+a_2$ over $F_2[x]/g(x)$ as
\begin{equation}\label{key311}
	\begin{aligned}	
		s=&a_2+a_1\\
		=&a_2+\frac{\big(Z_{i}-a_2(\beta_{2i})^2\big)-\big(Z_{j}-a_2(\beta_{2j})^2\big)}{\beta_{2i}-\beta_{2j}}.
	\end{aligned}
\end{equation}

Summing up the above two cases, we prove that any three participants can reconstruct the secret $s$. 

\noindent\textbf{The proof of Secrecy.} For any $C \in 2^{\mathcal{P}}\setminus\mathcal{A}$, we will prove that the secret $s$ is unable to be recovered by the shares of participants in $C$. $C$ is unqualified, then $|C|<3$. Similarly, we only prove the case of $|C|=2$. 

Denote the elements in $C$ as $P_{i+1}$ and $P_{j+1}$ with $0\leq i<j\leq 2^{\ell m}-1$. We consider the corresponding secret shares $Z_{i}$ and $Z_{j}$. Only one of the two following cases holds.

\noindent\textbf{Case 1} For the two shares, one is in ${\mathcal{Z}}$, the other is $a_2$.

\noindent\textbf{Case 2} Both $Z_{i}$ and $Z_{j}$ are in ${\mathcal{Z}}$.

\noindent\textbf{The proof of Case 1} Without loss of generality, suppose $Z_{i}\in {\mathcal{Z}}$ and $Z_{j}=a_2$. For any $s$, we have
\begin{equation*}
	Z_i=a_0+a_1(\beta_{2i})+a_2(\beta_{2i})^2.	
\end{equation*}
As $a_0$ and $a_1$ are the random variables uniformly distributed in the finite field $F_{2^\ell}[y]/g_1(y)$, $Z_i$ is independent from $a_2(\beta_{2i})^2$, which makes $Z_i$ independent from $s$ since $a_2=a_1+s$ over $F_2[x]/g(x)$. Hence, $Z_i$ is uniformly random in $F_{2^\ell}[y]/g_1(y)$ for each selection of $s$.

\noindent\textbf{The proof of Case 2} For this case, we will use Lemma~\ref{lem1} to prove the security. To use the conclusion of Lemma~\ref{lem1}, we choose any two distinct secret $s_0, s_1$, let $Z^{(s_0)}_{k}$ and $Z^{(s_1)}_{k}$ be corresponding the $(k+1)$-th shares for $P_{k+1}$. For any $z_i, z_j\in F_{2^\ell}[y]/g_1(y)$, we need to prove that the following two probabilities are equal, i.e.
\begin{equation}\label{key320}
	P(\{Z^{(s_0)}_i=z_i, Z^{(s_0)}_j=z_j\})=P(\{Z^{(s_1)}_i=z_i, Z^{(s_1)}_j=z_j\}.
\end{equation}

For given
$(\beta_{2i}, z_i)$ and $(\beta_{2j}, z_j)$, we respectively analyze the values of
$P(\{Z^{(s_0)}_i=z_i, Z^{(s_0)}_j=z_j\})$ and $P(\{Z^{(s_1)}_i=z_i, Z^{(s_1)}_j=z_j\}$. Over the finite field $F_{2^\ell}[y]/g_1(y)$, we can get the following equations
\begin{equation}\label{key321}
	\begin{aligned}
		\left
		\{
		\begin{array}{c}
			z_i=F(\beta_{2i})=a_{0,0}+a_{0,1}\beta_{2i}+a_{0,2}(\beta_{2i})^2,\\
			z_j=F(\beta_{2j})=a_{0,0}+a_{0,1}\beta_{2j}+a_{0,2}(\beta_{2j})^2,\\
		\end{array} \right.
	\end{aligned}
\end{equation}
and 
\begin{equation}\label{key322}
	\begin{aligned}
		\left
		\{
		\begin{array}{c}
			z_i=F(\beta_{2i})=a_{1,0}+a_{1,1}\beta_{2i}+a_{1,2}(\beta_{2i})^2,\\
			z_j=F(\beta_{2j})=a_{1,0}+a_{1,1}\beta_{2j}+a_{1,2}(\beta_{2j})^2,\\
		\end{array} \right.
	\end{aligned}
\end{equation}
where $a_{0,2}=a_{0,1}+s_0$ and $a_{1,2}=a_{1,1}+s_1$, and $a_{i,0},a_{i,1}\in F_{2^\ell}[y]/g_1(y)$, $a_{i,2}\in F_2[x]/g(x)$ for $i\in\{0,1\}$. 

The space of the solution vector $(a_{i,0},a_{i,1},a_{i,2})$ can be regarded as $\{ F_{2^\ell}[y]/g_1(y)\}^{2}\oplus\{F_2[x]/g(x)\}$, and denote $\{ F_{2^\ell}[y]/g_1(y)\}^{2}\oplus\{F_2[x]/g(x)\}$ as $U$. The value of $P(\{Z^{(s_0)}_i=z_i, Z^{(s_0)}_j=z_j\})$ is equal to the ratio of the number of solution $(a_{0,0},a_{0,1},a_{0,2})$ of \eqref{key321} in the whole space $U$. Similarly, the value of $P(\{Z^{(s_1)}_i=z_i, Z^{(s_1)}_j=z_j\})$ is equal to the ratio of the number of solution $(a_{1,0},a_{1,1},a_{1,2})$ of \eqref{key322} in the whole space $U$. Therefore, to prove \eqref{key320} holding, we only need to prove that the number of solutions $(a_{0,0},a_{0,1},a_{0,2})$ of \eqref{key321} and the number of solutions $(a_{1,0},a_{1,1},a_{1,2})$ of \eqref{key322} is equal.

We fisrt analyze the number of solutions $(a_{0,0},a_{0,1},a_{0,2})$ of \eqref{key321}. For \eqref{key321}, subtracte the first equation from the second equation in \eqref{key321} and get 
\begin{equation}\label{key323}
	z_j-z_i=a_{0,1}(\beta_{2j}-\beta_{2i})+a_{0,2}(\beta_{2j}-\beta_{2i})(\beta_{2j}+\beta_{2i}).
\end{equation}
In the finite field $F_{2^\ell}[y]/g_1(y)$, dividing both sides of the equation by $\beta_{2j}-\beta_{2i}$, then we have 
\begin{equation}\label{key324}
	\frac{z_j-z_i}{\beta_{2j}-\beta_{2i}}=a_{0,1}+a_{0,2}(\beta_{2j}+\beta_{2i}).
\end{equation}
Let $a_{0,1}=a_{0,\ell}+a_{0,h}y$, then $a_{0,2}=a_{0,\ell}+s_0$, where $a_{0,\ell}\in F_{2^\ell}$ and $a_{0,h}\in F_{2^{\ell(m-1)}}$. Substituting  $a_{0,1}=a_{0,\ell}+a_{0,h}y$ and $a_{0,2}=a_{0,\ell}+s_0$ into the above equation, we can get 
\begin{align}\label{key325}
	\frac{z_j-z_i}{\beta_{2j}-\beta_{2i}}=&a_{0,1}+a_{0,2}(\beta_{2j}+\beta_{2i})\nonumber\\
	=&a_{0,\ell}(1+\beta_{2j}+\beta_{2i})+a_{0,h}y+s_0(\beta_{2j}+\beta_{2i}).
\end{align}
From (\ref{key325}), we can get
\begin{equation}\label{key326}
	\frac{z_j-z_i}{\beta_{2j}-\beta_{2i}}-s_0(\beta_{2j}+\beta_{2i})=a_{0,\ell}(1+\beta_{2j}+\beta_{2i})+a_{0,h}y.
\end{equation}
Since $a_{0,\ell}\in F_{2^\ell}$ and $a_{0,h}\in F_{2^{\ell(m-1)}}$, the solution space of \eqref{key326} is $\{ F_{2^\ell}\}\oplus\{F_{2^{\ell(m-1)}}\}$. Thus, the number of solutions $(a_{0,0},a_{0,1},a_{0,2})$ of \eqref{key321} is equal to the number of solutions $(a_{0,\ell},a_{0,h})$ of \eqref{key326}.

We use a similar method to analyze the number of solutions $(a_{1,0},a_{1,1},a_{1,2})$ of \eqref{key322}. Subtracting the first equation from the second equation in \eqref{key322}, we can further get 
\begin{equation}\label{key327}
	z_j-z_i=a_{1,1}(\beta_{2j}-\beta_{2i})+a_{1,2}(\beta_{2j}-\beta_{2i})(\beta_{2j}+\beta_{2i}).
\end{equation}
Considering the finite field $F_{2^\ell}[y]/g_1(y)$, dividing both sides of the equation by $\beta_{2j}-\beta_{2i}$, then we have 
\begin{equation}\label{key328}
	\frac{z_j-z_i}{\beta_{2j}-\beta_{2i}}=a_{1,1}+a_{1,2}(\beta_{2j}+\beta_{2i}).
\end{equation}
Let $a_{1,1}=a_{1,\ell}+a_{1,h}y$, then $a_{1,2}=a_{1,\ell}+s_1$, where $a_{1,\ell}\in F_{2^\ell}$ and $a_{1,h}\in F_{2^{\ell(m-1)}}$. We take these variable substitutions into (\ref{key328}) and simplify (\ref{key328}), then 
\begin{equation}\label{key329}
	\frac{z_j-z_i}{\beta_{2j}-\beta_{2i}}-s_1(\beta_{2j}+\beta_{2i})=a_{1,\ell}(1+\beta_{2j}+\beta_{2i})+a_{1,h}y.
\end{equation}
The whole solution space of \eqref{key329} is $\{F_{2^\ell}\}\oplus\{F_{2^{\ell(m-1)}}\}$, which is the same as the solution space of \eqref{key326}. The number of solutions $(a_{1,0},a_{1,1},a_{1,2})$ for \eqref{key322} is equal to the number of solutions $(a_{1,\ell},a_{1,h})$ for \eqref{key329}.

Therefore, to prove \eqref{key320} holding, our goal has shifted from proving that the numbers of solutions of \eqref{key321} and \eqref{key322} equal to proving that the numbers of solutions of  (\ref{key326}) and (\ref{key329}) equal. 

We prove the numbers of solutions of (\ref{key326}) and (\ref{key329}) equal by contradiction. Without losing generality, assume that the equation (\ref{key326}) has $N_0$ solutions and the equation (\ref{key329}) has $N_1$ solutions with $N_0> N_1$. Let $\mathbb{A}_0$ and $\mathbb{A}_1$ respectively be the solution sets of (\ref{key326}) and (\ref{key329}). 

We subtracte the equation of (\ref{key326}) from the equation of (\ref{key329}) and get 
\begin{equation}\label{key330}
	(s_0-s_1)(\beta_{2j}+\beta_{2i})=(a_{1,\ell}-a_{0,\ell})(1+\beta_{2j}+\beta_{2i})+(a_{1,h}-a_{0,h})y.
\end{equation}
Let ${\Delta}_s=s_0-s_1$, ${\Delta}_{a_\ell}=a_{1,\ell}-a_{0,\ell}$, and ${\Delta}_{a_h}=a_{1,h}-a_{0,h}$, where ${\Delta}_s,{\Delta}_{a_\ell}\in F_{2^\ell} $ and ${\Delta}_{a_h}\in F_{2^{\ell(m-1)}}$. Substitute the above results into (\ref{key330}) and simplify it as
\begin{equation}\label{key331}
	{\Delta}_s(\beta_{2j}+\beta_{2i})={\Delta}_{a_\ell}(1+\beta_{2j}+\beta_{2i})+{\Delta}_{a_h}y.
\end{equation}
Solving the above equation, we can construct a solution of \eqref{key331}, i.e.
\begin{equation}
	{\Delta}_{a_\ell}={\Delta}_s(\beta_{2j}+\beta_{2i})(1+\beta_{2j}+\beta_{2i})^{-1} \quad (\textrm{over}\ F_{2^\ell})
\end{equation}
and 
\begin{equation}
	{\Delta}_{a_h}=\frac{{\Delta}_s(\beta_{2j}+\beta_{2i})-{\Delta}_{a_\ell}(1+\beta_{2j}+\beta_{2i})}{y}\ (\textrm{over}\ F_{2^{\ell(m-1)}}),
\end{equation}
we denote the solution as ${\Delta}^{*}=({\Delta}^{*}_{a_\ell}, {\Delta}^{*}_{a_h})$.

According to the assumptions, the equation (\ref{key326}) has solutions. As $\mathbb{A}_0$ is the solution set of the equation (\ref{key326}), we choose one of solutions in $\mathbb{A}_0$ and denote it as $a^1_0=(a^1_{0,\ell}, a^1_{0,h})$, then $a^1_0+{\Delta}^{*}=(a^1_{0,\ell}+{\Delta}^{*}_{a_\ell}, a^1_{0,h}+{\Delta}^{*}_{a_h})$ is a solution of the equation (\ref{key329}) by suming two equations (\ref{key326}) and (\ref{key330}). When $a^1_{0}$ is taken across the whole solution set $\mathbb{A}_0$, the
equation (\ref{key329}) has at least $N_0$ different solutions, which contradicts that (\ref{key329}) has $N_1$ solutions. Therefore, the equations (\ref{key326}) and (\ref{key329}) have the same number of solutions. 

\subsection{The Construction of Evolving $3$-threshold Scheme}
Now, we choose the evolving $3$-threshold scheme in~\cite{cheng2024construction} as $\mathcal{S}^{\infty}$ and the $3$-threshold scheme mentioned in Section~\ref{sec5.1} as $\mathcal{S}^{i}$. The correctness and secrecy of the two schemes have been proved, which are shown in~\cite{cheng2024construction} and Section\ref{sec5.1}. Therefore, the correctness and secrecy of the new construction of the evolving $3$-threshold secret sharing scheme are perfect, the corresponding proofs are described in Section~\ref{sub4.2}.

Similarly, for the $t$-th participant, it belongs to the $g(t)=\lceil \log_4 \lg t\rceil$-th generation, thus the number of participants for the $g(t)$-th generation is $S(G^{g(t)})$. The corresponding shares of the $t$-th participant is composed of several pieces, which have been shown in Section~\ref{sub4.1}. Specifically, for the $3$-threshold scheme $\mathcal{S}^{g(t)}$, it needs to constructs $S(G^{g(t)})+1$ shares. In Section~\ref{sub5.1}, $\mathcal{S}^{i}$ scheme can constrcuts the shares $\{F(2i)\}_{i=0}^{2^{\ell m-1}-1}\cup\{a_2\}$. Thus, 
\begin{equation}\label{key601}
	2^{\ell m-1}+1\geq S(G^{g(t)})+1,
\end{equation}
solving the above equation, we get $m\geq \frac{ \lg S(G^{g(t)})+1}{\ell}$.

Among $2^{\ell m-1}+1$ shares, the share $F(2(h(t)-1))$ is distributed to the $t$-th participant, where $h(t)$ denotes the index of $t$-th participant in the $g(t)$-th generation with $S(G^{g(t)})$ participants. The share $a_2$ is regarded as the forward share ${sh}_{F}^{g(t)}$ with $\ell$-bit length.
\begin{table*}[t]
	\renewcommand\arraystretch{1}
	\centering
	\caption{Share sizes of evolving $k$-threshold schemes.}
	\setlength{\tabcolsep}{1.5mm}
	\begin{tabular}{|c|c|c|}
		\hline
		threshold &algorithm &share size\\\hline
		\multirow{3}*{$k=3$}
		&\cite{komargodski2017share}&$2\lg t+486\ell\lg{\lg t}\cdot\lg{\lg {\lg t}}+ 567\ell\lg 3$\\
		\cline{2-3}
		~&\cite{cheng2024construction}& $2 \lfloor\lg t\rfloor+4\lfloor\lg ({\lfloor\lg t\rfloor+1}) \rfloor+\ell$\\
		\cline{2-3}
		~& \textbf{ours}& \bm {${\lg t+O({\lceil \log_4 \lg t\rceil}^2)+\ell(2\lceil \log_4 \lg t\rceil-1)+1}$} \\		
		\hline				
	\end{tabular}
	\label{tab1}
\end{table*} 

\subsection{Share Size}
The shares of the $t$-th participant are composed of five parts, $P1$, $P2$, $P3$, $P4$ and $P5$ for $g(t)\geq 2$. Therefore, $Z^{(s)}_{P_t}$ is the sum of share size of all shares in $P1$, $P2$, $P3$, $P4$ and $P5$. And we denote the share size of all shares in $Pi$ as $B(Pi)$, where $1\leq i\leq 5$. Then, we have
\begin{equation}\label{key602}
	B(Z^{(s)}_{P_t})=B(P1)+B(P2)+B(P3)+B(P4)+B(P5).
\end{equation}
Observing (\ref{key602}), $B(P1)$ is equal to $B({sh}_{g(t)}^{\infty})$ with the share size is equal to $O({g(t)}^2)$, which is mentioned in~\cite{cheng2024construction}. Since $g(t)=\lceil \log_4 \lg t\rceil$, then
\begin{equation}\label{key603}
	B(P1)=B({sh}_{g(t)}^{\infty})=O({g(t)}^2)=O({\lceil \log_4 \lg t\rceil}^2).
\end{equation}
And the bit length of $B(P2)+B(P4)+B(P5)$ is shown in \eqref{key433}, combining ${sh}_{F}^{g(t)}$ is $\ell$-bit length, then we have
\begin{equation}\label{key604}
	\begin{aligned}
		&B(P2)+B(P4)+B(P5)\\
		=&\sum_{j=1}^{g(t)-1}B({sh}_{F}^{j})
		+\sum_{j=1}^{g(t)-1}B({SR}^{j}\oplus {sh}_{F}^{g(t)})+B({SR}^{g(t)})\\
		=&\big(2g(t)-1\big)B({sh}_{F}^{1})\\
		=&\ell(2\lceil \log_4 \lg t\rceil-1).
	\end{aligned}
\end{equation}

As for $B(P3)$, it is equal to $B({sh}_{h(t)}^{g(t)})$. The scheme $\mathcal{S}^{g(t)}$ construct the share ${sh}_{h(t)}^{g(t)}$ in the field field $F_{2^{\ell m}}$, hence, the share size is no more than $\ell m$. Since $m\geq \frac{ \lg S(G^{g(t)})+1}{\ell}$, then $\ell m\geq \lg S(G^{g(t)})+1$.

Combining the values of $B(P1)$, $B(P2)+B(P4)+B(P5)$ and $B(P3)$, we obtain
\begin{equation}\label{key605}
	\begin{aligned}
		B(Z^{(s)}_{P_t})=&B(P1)+B(P2)+B(P4)+B(P5)+B(P3)\\
		=&O({\lceil \log_4 \lg t\rceil}^2)+\ell(2\lceil \log_4 \lg t\rceil-1)\\
		&+\lg S(G^{g(t)})+1\\
		\overset{(a)}\leq&\lg t+O({\lceil \log_4 \lg t\rceil}^2)+\ell(2\lceil \log_4 \lg t\rceil-1)+1,	
	\end{aligned}	
\end{equation}
where ($a$) holds since $S(G^{g(t)})\leq t$.

\subsection{Comparision}
In this subsection, we tabulate the share sizes of currently known evolving $3$-threshold secret sharing schemes. In Table~\ref{tab1}, we represent the value of the lowest share size in bold. For the scheme~\cite{cheng2024construction}, we show the corresponding share size for using $\delta$ code as the prefix code. By comparing, we find that the most important term is $2\lg t$ for the schemes in \cite{komargodski2017share,cheng2024construction}, and the most important term is only $\lg t$ for the proposed scheme, which can achieve consistently smaller share sizes than the schemes in \cite{komargodski2017share,cheng2024construction}.

\section{Conclusion and Discussion}\label{sec6}
In this paper, for the known evolving $3$-threshold secret sharing scheme mentioned in~\cite{DARCO2021149}, we have identified security issues. Based on this, we propose an evolving $3$-threshold scheme for an $\ell$-bit secret with perfect security and smaller share size, the result improves the prior best result. Besides, we also propose a conventional $3$-threshold secret sharing scheme over a finite field. Based on this model of the revised scheme, then we present a brand-new evolving $3$-threshold secret sharing scheme. There are still some thought-provoking and challenging issues that remain unresolved, which are also our future works.

%(1) The correctness and security of the proposed schemes are perfect. If relaxing correctness or security, is it possible to propose more interesting and efficient schemes to achieve better efficiency or smaller share size?

(1) The share size of the proposed evolving $3$-threshold scheme is uncertain whether it is optimal. If so, please explain. If not, whether the evolving $3$-threshold scheme proposed to achieve a smaller share size?

(2) The proposed scheme is only applicable to the case of $k=3$, is it possible to extend it to a more general threshold $k$ to improve the smaller share size than \cite{cheng2024construction}, where $k\geq 4$.

\bibliographystyle{IEEEtran} 
\bibliography{IEEEabrv,refs}

% Generated by IEEEtran.bst, version: 1.14 (2015/08/26)
\begin{thebibliography}{10}
\providecommand{\url}[1]{#1}
\csname url@samestyle\endcsname
\providecommand{\newblock}{\relax}
\providecommand{\bibinfo}[2]{#2}
\providecommand{\BIBentrySTDinterwordspacing}{\spaceskip=0pt\relax}
\providecommand{\BIBentryALTinterwordstretchfactor}{4}
\providecommand{\BIBentryALTinterwordspacing}{\spaceskip=\fontdimen2\font plus
\BIBentryALTinterwordstretchfactor\fontdimen3\font minus
  \fontdimen4\font\relax}
\providecommand{\BIBforeignlanguage}[2]{{%
\expandafter\ifx\csname l@#1\endcsname\relax
\typeout{** WARNING: IEEEtran.bst: No hyphenation pattern has been}%
\typeout{** loaded for the language `#1'. Using the pattern for}%
\typeout{** the default language instead.}%
\else
\language=\csname l@#1\endcsname
\fi
#2}}
\providecommand{\BIBdecl}{\relax}
\BIBdecl

\bibitem{shamir1979share}
A.~Shamir, ``How to share a secret,'' \emph{Communications of the ACM},
  vol.~22, no.~11, pp. 612--613, Nov. 1979.

\bibitem{blakley1979safeguarding}
G.~R. Blakley, ``Safeguarding cryptographic keys,'' in \emph{Managing
  Requirements Knowledge, International Workshop on}.\hskip 1em plus 0.5em
  minus 0.4em\relax IEEE Computer Society, 1979, pp. 313--313.

\bibitem{fuyou2014randomized}
M.~Fuyou, X.~Yan, W.~Xingfu, and M.~Badawy, ``Randomized component and its
  application to ($ t $, $ m $, $ n $)-group oriented secret sharing,''
  \emph{IEEE Transactions on Information Forensics and Security}, vol.~10,
  no.~5, pp. 889--899, 2014.

\bibitem{harn2016realizing}
L.~Harn, C.~Hsu, M.~Zhang, T.~He, and M.~Zhang, ``Realizing secret sharing with
  general access structure,'' \emph{Information Sciences}, vol. 367, pp.
  209--220, 2016.

\bibitem{harn2010strong}
L.~Harn and C.~Lin, ``Strong (n, t, n) verifiable secret sharing scheme,''
  \emph{Information Sciences}, vol. 180, no.~16, pp. 3059--3064, 2010.

\bibitem{pang2005new}
L.-J. Pang and Y.-M. Wang, ``A new (t, n) multi-secret sharing scheme based on
  shamir’s secret sharing,'' \emph{Applied Mathematics and Computation}, vol.
  167, no.~2, pp. 840--848, 2005.

\bibitem{yang2004t}
C.-C. Yang, T.-Y. Chang, and M.-S. Hwang, ``A (t, n) multi-secret sharing
  scheme,'' \emph{Applied Mathematics and Computation}, vol. 151, no.~2, pp.
  483--490, 2004.

\bibitem{csirmaz2012line}
L.~Csirmaz and G.~Tardos, ``On-line secret sharing,'' \emph{Designs, Codes and
  Cryptography}, vol.~63, pp. 127--147, 2012.

\bibitem{komargodski2016share}
I.~Komargodski, M.~Naor, and E.~Yogev, ``How to share a secret, infinitely,''
  in \emph{Theory of Cryptography Conference}.\hskip 1em plus 0.5em minus
  0.4em\relax Springer, 2016, pp. 485--514.

\bibitem{komargodski2017share}
{I. Komargodski}, M.~Naor, and E.~Yogev, ``How to share a secret, infinitely,''
  \emph{IEEE Transactions on Information Theory}, vol.~64, no.~6, pp.
  4179--4190, Jun. 2017.

\bibitem{komargodski2017evolving}
I.~Komargodski and A.~Paskin-Cherniavsky, ``Evolving secret sharing: dynamic
  thresholds and robustness,'' in \emph{Theory of Cryptography: 15th
  International Conference, TCC 2017, Baltimore, MD, USA, November 12-15, 2017,
  Proceedings, Part II 15}.\hskip 1em plus 0.5em minus 0.4em\relax Springer,
  2017, pp. 379--393.

\bibitem{dutta2019secret}
S.~Dutta, P.~S. Roy, K.~Fukushima, S.~Kiyomoto, and K.~Sakurai, ``Secret
  sharing on evolving multi-level access structure,'' in \emph{International
  Workshop on Information Security Applications}.\hskip 1em plus 0.5em minus
  0.4em\relax Springer, 2019, pp. 180--191.

\bibitem{d2018equivalence}
P.~D’Arco, R.~D. Prisco, and A.~D. Santis, ``On the equivalence of
  2-threshold secret sharing schemes and prefix codes,'' in \emph{International
  Symposium on Cyberspace Safety and Security}.\hskip 1em plus 0.5em minus
  0.4em\relax Springer, 2018, pp. 157--167.

\bibitem{okamura2020new}
R.~Okamura and H.~Koga, ``New constructions of an evolving 2-threshold scheme
  based on binary or d-ary prefix codes,'' in \emph{2020 International
  Symposium on Information Theory and Its Applications (ISITA)}.\hskip 1em plus
  0.5em minus 0.4em\relax IEEE, 2020, pp. 432--436.

\bibitem{cheng2024construction}
Q.~Cheng, H.~Cao, S.-J. Lin, and N.~Yu, ``A construction of evolving $ k
  $-threshold secret sharing scheme over a polynomial ring,'' \emph{arXiv
  preprint arXiv:2402.01144}, 2024.

\bibitem{elias1975universal}
P.~Elias, ``Universal codeword sets and representations of the integers,''
  \emph{IEEE transactions on information theory}, vol.~21, no.~2, pp. 194--203,
  1975.

\bibitem{DARCO2021149}
P.~D'Arco, R.~{De Prisco}, and A.~{De Santis}, ``Secret sharing schemes for
  infinite sets of participants: A new design technique,'' \emph{Theoretical
  Computer Science}, vol. 859, pp. 149--161, 2021.

\end{thebibliography}
\end{document}